\documentclass[twocolumn]{aastex631}
\usepackage{mathrsfs}
\usepackage{lineno}

\newcommand{\alphaCO}{\alpha_{\rm CO}}
\newcommand{\hi}{{\rm H}{\textsc i}}
\newcommand{\rhi}{R_{\rm HI}}
\newcommand{\rnine}{{r}_{\rm 90}}
\newcommand{\rfive}{{r}_{\rm 50}}
\newcommand{\dhi}{D_{\rm HI}}
\newcommand{\mhi}{M_{\rm HI}}
\newcommand{\mstar}{M_{\rm *}}
\newcommand{\mhiin}{M_{\rm HI,in}}
\newcommand{\mhiout}{M_{\rm HI,out}}

\newcommand{\htwo}{{\rm H}_2}
\newcommand{\mdust}{M_{\rm{dust}}}
\newcommand{\mgas}{M_{\rm{gas}}}
\newcommand{\mgasin}{\it{M}_{\rm{gas,in}}}
\newcommand{\mHtwo}{M_{{\rm H_2}}}
\newcommand{\msun}{M_{{\rm \odot}}}
\newcommand{\xihi}{\xi_{{\rm HI}}}
\newcommand{\xihiin}{\xi_{{\rm HI,in}}}
\newcommand{\xihtwo}{\xi_{{\rm H_2}}}
\newcommand{\xigas}{\xi_{{\rm gas}}}
\newcommand{\xigasin}{\xi_{{\rm gas,in}}}

\newcommand{\lsun}{L_{{\rm \odot}}}
\newcommand{\LWthree}{L_{\rm 12\mu m}}
\newcommand{\Lfive}{L_{\rm 500\mu m}}
\newcommand{\Ltwofive}{L_{\rm 250\mu m}}
\newcommand{\Lsixteen}{L_{\rm 160\mu m}}
\newcommand{\LTIR}{L_{\rm TIR}}
\newcommand{\LWthreeMdust}{L_{\rm 12\ \mu m}/M_{\rm dust}}
\newcommand{\MdustMstar}{M_{\rm dust}/M_{\rm *}}
\newcommand{\LNII}{L_{\rm [NII]}}
\newcommand{\Nii}{[\rm N\textsc{ii}]}
\newcommand{\cii}{[\rm C\textsc{ii}]}
\newcommand{\oii}{[\rm O\textsc{ii}]}

\newcommand{\oiii}{[\rm O\textsc{iii}]}

\shortauthors{}
\graphicspath{{./}{}}

\begin{document}

\title{The ISM scaling relations using inner HI and an application of estimating dust mass }
\correspondingauthor{Jing Wang}
\email{jwang\_astro@pku.edu.cn}

\author[0000-0001-9472-2052]{Fujia Li}
\affiliation{Deep Space Exploration Laboratory / Department of Astronomy, University of Science and Technology of China, Hefei 230026, China}
\affiliation{School of Astronomy and Space Science, University of Science and Technology of China, Hefei 230026, China}

\author[0000-0002-6593-8820]{Jing Wang}
\affiliation{Kavli Institute for Astronomy and Astrophysics, Peking University, Beijing 100871, China}

\author[0000-0001-5950-1932]{Fengwei Xu}
\affiliation{Kavli Institute for Astronomy and Astrophysics, Peking University, Beijing 100871, China}
\affiliation{Department of Astronomy, School of Physics, Peking University, Beijing 100871, People’s Republic of China}

\author[0000-0002-7660-2273]{Xu Kong}
\affiliation{Deep Space Exploration Laboratory / Department of Astronomy, University of Science and Technology of China, Hefei 230026, China}
\affiliation{School of Astronomy and Space Science, University of Science and Technology of China, Hefei 230026, China}

\author[0000-0002-5016-6901]{Xinkai Chen}
\affiliation{Kavli Institute for Astronomy and Astrophysics, Peking University, Beijing 100871, China}
\affiliation{Deep Space Exploration Laboratory / Department of Astronomy, University of Science and Technology of China, Hefei 230026, China}
\affiliation{School of Astronomy and Space Science, University of Science and Technology of China, Hefei 230026, China}

\author[0000-0001-8078-3428]{Zesen Lin}
\affiliation{Department of Physics, The Chinese University of Hong Kong, Shatin, N.T., Hong Kong S.A.R., China}
\affiliation{Deep Space Exploration Laboratory / Department of Astronomy, University of Science and Technology of China, Hefei 230026, China}
\affiliation{School of Astronomy and Space Science, University of Science and Technology of China, Hefei 230026, China}

\author[0000-0002-9663-3384]{Shun Wang}
\affiliation{Kavli Institute for Astronomy and Astrophysics, Peking University, Beijing 100871, China}
\affiliation{Department of Astronomy, School of Physics, Peking University, Beijing 100871, People’s Republic of China}

\begin{abstract}
We select a disk-like galaxy sample with observations of the $\hi$, $\htwo$ and dust from {\it Herschel} Reference Survey (HRS), and derive inner $\hi$ masses within the optical radius. We find that the inner gas-to-dust ratio is almost independent of gas-phase metallicity, and confirm that the inner gas mass ($\hi$+$\htwo$) shows tighter relationship with dust mass and monochromatic 500 $\mu m$ luminosity than the integral gas mass. It supports that dust is more closely associated with co-spatial cold gas than the overall cold gas. Based on the newly calibrated relationship between inner gas mass and dust mass, we predict dust masses for disk-dominated galaxies from the xCOLD GASS sample. The predicted dust masses show scaling relations consistent with fiducial ones in the literature, supporting their robustness. Additionally, we find that at a given dust mass and star formation rate (SFR), the galactic WISE W3 luminosities show significant dependence on the $\Nii$ luminosity and the stellar mass surface density. Such dependence highlights the caveat of using the W3 luminosity as integral SFR indicator, and is consistent with findings of studies which target star-forming regions in more nearby galaxies and accurately derive dust masses based on mapping-mode spectroscopy. 
\end{abstract}


\section{Introduction}

A comprehensive census of multi-phase interstellar medium (ISM) and its hydrodynamic processes can provide important insights into the understanding of galaxy evolution, despite the difficulties to observe them. 

At the present, gathering statistically large samples of ISM (i.e., the atomic neutral hydrogen gas $\hi$, the molecular hydrogen gas $\htwo$, and the dust) properties remains challenging. 
For $\hi$ gas, either the sample sizes are small and biased when the data are collected with interferometry, or the spatial resolution of data are low and only integral spectra can be obtained with single-dish telescopes \citep{Saintonge2022}. The situation is improved with SKA pathfinder surveys conducted at the Australian Square Kilometre Array Pathfinder \citep[ASKAP;][]{Johnston2007,Johnston2008} and the South African Meer-Karoo Array Telescope \citep[MeerKAT;][]{Jonas2016}. But the majority of galaxies detected in these new surveys remain unresolved or marginally resolved \citep{Koribalski2020,Westmeier2022}. 
The studies of $\htwo$ are usually conducted through observing the CO emission, as done by the extended CO Legacy Database for GASS \citep[xCOLD GASS; ][]{Saintonge2011, Saintonge2017}. CO observations have reached a higher redshift than $\hi$ \citep[e.g.,][]{Daddi2010,Aravena2014,Riechers2020,Kaur2022}, despite the uncertainty of conversion between CO and $\htwo$ masses related to the metallicity, pressure, and radiation field \citep[e.g.,][]{Schruba2012,Accurso2017}. 
Dust is typically observed in the infrared (IR) as dust grains absorb the light from stars and re-emit across the entire IR band.
Dust mass is dominated by large dust grains in size, but small dust grains provide useful diagnostics of ISM properties \citep{Draine2007}. 
Most of the observations have been conducted by space telescopes, such as IRAS \citep{Neugebauer1984}, Spitzer \citep{Werner2004}, and {\it Herschel} \citep{Pilbratt2010}, and are thus expensive. 

In spite of these difficulties, studies based on existing observations have shown a dynamic picture of ISM evolution and its link to galaxy evolution. Through the process of star formation, the different components of ISM ($\hi$, $\htwo$, and dust) are tightly connected to each other. Dust can act as a catalyst in the transformation from $\hi$ to $\htwo$ gas and also plays an important role in balancing the heating and cooling processes of gas in galaxy \citep{Hollenbach1971,Hollenbach1997}. $\htwo$ further condenses and forms stars, which when evolve return the ISM with dust and gas \citep{Kennicutt1998,Krumholz2012}. 

The trend of gas-to-dust ratio with metallicity is an important scaling relation to constrain dust evolution models and to link the $\hi$, $\htwo$, and dust \citep{Draine2007,Galliano2018}. For high-redshift galaxies, it has become an increasingly common method to estimate total gas (i.e., $\hi$ and $\htwo$ together) mass (gas mass for short) or ISM mass via the gas-to-dust ratio calibrated in the local Universe \citep{Leroy2011,Remy-Ruyer2014,Janowiecki2018}. A tricky issue involved in the gas-to-dust ratio, is the different radial extensions of multi-phase gas, dust, and metallicity \citep[e.g.,][]{Leroy2008,MuozMateos2009,Abdurro'uf2022}. For gas-rich star-forming galaxies, the typically measured metallicity and $\htwo$ gas are almost within the optical stellar disks, whereas $\hi$ disks can be much more extended . Previous statistics suggest that the typical ratio between the sizes of $\hi$ and optical disks is between 2 and 4, with an average  value of 1.7 \citep[e.g.,][]{Broeils1997,Hunter1997,Swaters2002,Thuan2004,Wang2013}. In the outer disk, metallicity decreases with radius, whereas the surface density ratios of $\Sigma_{\rm gas} / \Sigma_{\rm dust}$ and $\Sigma_{\hi} / \Sigma_{\htwo}$ show the opposite behavior \citep{Alton1998,Bianchi2007,Bigiel2008,MuozMateos2009,Hunt2015,Belfiore2017,Casasola2017}. So, when investigating the relation between the integral values of $\mgas$, $\mdust$, and metallicity as commonly done in the literature, the metallicity tends to be over-estimated if the region in question is assumed to be the whole gas disk, whereas $\mgas / \mdust$ tends to be over-estimated if the region assumed to be the stellar disk. As a result, the steepness of the relation between gas-to-dust ratio and metallicity can be over-estimated. A difficulty in solving this bias comes from the limited number of spatially resolved $\hi$ images for galaxies. \citet{Wang2020} have therefore introduced a useful method to estimate from integral $\hi$ mass the inner part of $\hi$ mass within optical radius of disk-like galaxies, based on statistical properties of $\hi$ disks. In this paper, we use the method of \citet{Wang2020} to obtain the inner $\hi$ mass and calibrate a new dust--inner gas mass relation, which shows a stronger and tighter correlation than the dust--integral gas mass relation.

As a first application and test of the newly calibrated dust--inner gas mass relation, we look into the goodness of the mid infrared (MIR) W3 12 $\mu \rm m$ band from the Wide-field Infrared Explorer \citep[WISE;][]{Wright2010} as the star formation rate (SFR) indicator. The monochromatic MIR luminosities can be used to indicate SFR because of the facts that polycyclic aromatic hydrocarbons (PAHs) and small warm dust grains can be heated in the photodissociation regions (PDRs) \citep[e.g.,][]{Popescu2000, Wu2005, Rieke2009,Boquien2021}. However, PAHs can be affected by the strong and hard radiation fields associated with low-metallicity systems \citep{Smith2007, Aniano2020}, leading to a deficit in the PAHs luminosity. \citet{Gregg2022} find a strong metal-dependent deviation in the relation of SFR surface density and 8 $\mu \rm m$ surface brightness when focusing on the brightest PAHs emission peaks. On the other hand, when combining the {\it Spitzer} Infrared Nearby Galaxy Survey \citep[SINGS;][]{Kennicutt2003} and the Key Insights on Nearby Galaxies: a Far-infrared Survey with {\it Herschel} \citep[KINGFISH;][]{Kennicutt2011} data with the WISE data to calibrate the MIR W3 12 $\mu \rm m$ SFR indicator, \citet{Cluver2017} find insignificant metal-dependence in the deviations between the total IR (TIR) and the 12 $\mu \rm m$ indicated SFRs. It is possible, as \citet{Cluver2017} has extensively discussed that, the contributors to the relatively broad W3 band are more complex than that of the 8 $\mu \rm m$, but it is also possible that the sample used is relatively small or biased. 
We thus take a new look into the influence of galactic internal environments, including the metallicity, ionization parameter, and stellar mass surface density of old stars, on the W3 12 $\mu \rm m$ luminosity as an SFR indicator, by using a larger and relatively representative sample of disk galaxies. We estimate the dust mass of xCOLD GASS sample via our newly calibrated inner gas--dust mass relation and use it to study the metallicity dependence of 12 $\mu \rm m$ at given $\mdust$ and SFR. 

The paper is organized as follows. We introduce and describe our sample selection from the {\it Herschel} Reference Survey \citep[HRS, ][]{Boselli2010}, xCOLD GASS, and the JCMT dust and gas In Nearby Galaxies Legacy Exploration \citep[JINGLE][]{Saintonge2018} in Section \ref{sec:sample}. In Section \ref{sec:gas-to-dust ratio} and Section \ref{sec:relation}, we use inner $\hi$ and inner gas to calibrate new ISM scaling relations. In Section \ref{sec:dust mass}, we use inner gas--dust mass relation to estimate dust mass of xCOLD GASS sample and compare it to JINGLE and Section \ref{sec:L12/Mdust} includes discussions of the WISE W3 SFR indicator using $\LWthree$/$\mdust$. We discuss the applications of the inner gas in Section \ref{sec:discussion}. Finally, we briefly summarize our main results in Section \ref{sec:summary}. Throughout this work, we adopt a $\Lambda$ cold dark matter cosmology with $\Omega_{m}=0.3$, $\Omega_{\lambda}=0.7$ and $h=0.7$.

\section{sample and data}\label{sec:sample}
We build three samples in the following, the HRS disk sample, the xCOLD GASS disk sample, and the JINGLES disk sample. The first sample will be used to calibrate a relation between dust mass and inner gas mass. We will apply the relation to the second sample to predict dust masses from inner gas masses, and investigate relationship between the 12 $\mu$m luminosity, predicted dust mass, and SFR. The third sample will mainly be used for comparison and show the less biased galactic population of the second sample. In Table \ref{tab:data} we have collected the median values of the three disk sample properties.

\subsection{Herschel Reference Survey}
The HRS sample of 322 galaxies\footnote{With respect to the original sample given in \citet{Boselli2010}, the galaxy HRS 228 is removed from the complete sample because its updated redshift on NED indicates it as a background object.} is an IR survey ($100-500\ \mu \rm m$) with multi-wavelength data set and selected by the K-band (2.2 $\mu \rm m$) magnitude and luminosity distance (between 15 and 25 Mpc). We use data from HRS to calibrate the relation between dust mass and inner gas mass.

Following \citet{Janowiecki2018}, we select the galaxies with measurements of the atomic gas mass $\mhi$, the molecular gas mass $\mHtwo$ (we also take into account the CO non-detection), each monochromatic infrared fluxes in {\it Herschel} bands, the dust mass $\mdust$, and the gas-phase metallicity $\rm 12+\log (O/H)$. We select ``$\hi$-normal" galaxies that have $\hi$ deficiency parameter less than or equal to 0.5. Furthermore, we require $r$-band optical concentration $\rnine/\rfive <2.7$ to select relatively disk dominated galaxies (see Section \ref{subsec:inner HI}). Finally, these selections result in a sample of $\rm N = 55$ galaxies (42 CO detected galaxies and 13 non-detections). Note that all galaxies are $\hi$ detections in our selections. This sample is referred to as the HRS disk sample.

The difference in sample size from that of \citet{Janowiecki2018} is partly due to the different choice of gas-phase metallicity (which removes 13 galaxies, see Section \ref{subsec:HRS selection}), and also due to the cut in $r$-band optical concentration ($\rnine/\rfive <2.7$, which removes 12 galaxies). 

\subsubsection{Galaxy properties of the HRS disk sample}\label{subsec:HRS selection}
We use the following properties of galaxies from the HRS disk sample for analysis.  

$\hi$ gas mass ($\mhi$): We use the atomic gas mass ($\mhi$, in $\msun$) from \citet{Boselli2014}. They collected the $\hi$ integrated spectra from a wide range of literature references for 315 out of the 322 HRS galaxies. If the rms of $\hi$ spectra are available, the flux errors are propagated into $\mhi$ uncertainties; otherwise, a typical uncertainty of 15 percent in the mass is considered.

Molecular hydrogen mass ($\mHtwo$): We use the molecular hydrogen mass estimated from the CO(1-0) observations which are a compilation of archival data in \citet{Boselli2014}. Different from the conversion factor $\alpha_{CO}$ used in HRS sample \citep{Boselli2002}, which depends on the H band luminosity, we use the function calibrated by \citet{Accurso2017}. They corrected for the contaminant $\cii$ emission and used the $L_{\cii}/L_{\rm{CO(1-0)}}$ ratio as well as radiative transfer modelling to introduce a new conversion factor, which depends on the metallicity and the offset from the star-forming main sequence $(\Delta (\mbox{MS}))$. The flux error includes the measurement uncertainty of CO(1-0) line luminosity and the 35\% uncertainty on the $\alphaCO$ conversion factor \citep{Accurso2017,Saintonge2017}. We note that using the constant or K-band based $\alphaCO$ causes small changes in slopes and intercepts of relations, but does not affect the major trends or conclusions in this paper.

Integral gas mass ($\mgas$): Throughout this work, as a comparison with inner gas mass (see Section \ref{subsec:inner HI}), we define integral gas mass, which includes the integral $\hi$ and $\htwo$ masses as well as a 36\% correction to account for heavy elements: $\mgas = 1.36 \times (\mhi + \mHtwo)$. The uncertainties are propagated from $\mhi$ and $\mHtwo$.

The 500 $\mu \rm m$ luminosity ($\Lfive$) and dust mass ($\mdust$): The monochromatic IR data in HRS are obtained from the Photoconductor Array Camera and Spectrometer (PACS; 100 and 160 $\mu \rm m$) and the Spectral and Photometric Imaging Receiver (SPIRE; 250, 350 and 500 $\mu \rm m$) instruments. We use the 500 $\mu \rm m$ fluxes obtained from \citet{Ciesla2012} and convert them to luminosity in solar units ($\lsun$). The dust masses come from \citet{Ciesla2014}, which obtain them by fitting the spectral energy distributions (SED) between 8 and 500 $\mu \rm m$ with the dust models of \citet{Draine2007}. The uncertainties are also given by \citet{Ciesla2014}. 

Gas phase metallicity ($\rm 12+\log (O/H)$): \citet{Hughes2013} used integrated, drift-scan optical spectroscopic observations to obtain reliable metallicity estimates for the HRS. They estimated oxygen abundances with five different calibration methods, and also calculate the average of them. When all metallicities are derived with the O3N2 calibration from \citet{Pettini2004}(hereafter PP04), the results show small systematic discrepancies from other methods \citep{Kewley2008,Hughes2013,DeVis2019,Maiolino2019,Yates2021}. In addition, the metallicity used to derive the CO conversion factor $\alphaCO$ is in the PP04 type. So in this work, we use the metallicity derived with the PP04 O3N2 method. The uncertainties are also provided by \citet{Hughes2013}.

Stellar mass ($M_{*}$) and star formation rate (SFR): $M_{*}$ and SFR are used in the calculation of the conversion factor $\alphaCO$. We use SFRs from \citet{Boselli2015}, which combines 4 different estimates from $H\alpha$+Balmer decrement, $H\alpha$+24$\mu m$, FUV+24$\mu m$, and radio 20cm emission. The stellar masses $M_{*}$ come from \citet{Cortese2012} and are estimated from $i$-band luminosities $L_{i}$ using the ($g$-$i$)-dependent stellar mass-to-light ratio from \citet{Zibetti2009}, assuming a \citet{Chabrier2003} initial mass function (IMF). 

Optical $r$-band radius ($\rfive$ and $\rnine$): $\rfive$ and $\rnine$ are the radii enclosing 50\% and 90\% of the total flux in optical $r$-band. The radii are be taken from the NASA Sloan Atlas \citep[NSA; ][]{Blanton2011} and will be used to estimate the inner gas mass (see Section \ref{subsec:inner HI}).

\subsection{xCOLD GASS}\label{subsec:xCOLD GASS}
The extended GALEX Arecibo SDSS Survey \citep[xGASS; ][]{Catinella2018} selected galaxies by stellar mass ($10^{9}\leq M_{*}/M_{\odot}\leq 10^{11.5}$) and redshift ($0.01<z<0.05$) from the SDSS DR7 \citep{Abazajian2009} spectroscopy survey and GALEX \citep{Martin2005} imaging survey. The $\hi$ data of 1179 galaxies were taken with Arecibo telescope and complemented by data from the Arecibo Legacy Fast ALFA \citep[ALFALFA;][]{Giovanelli2005, Haynes2011} and Cornell $\hi$ digital archive \citep{Springob2005}. The xCOLD GASS \citep{Saintonge2011} targeted an unbiased subsample of 532 xGASS galaxies with IRAM 30m CO(1-0) observations. The main advantages of these surveys lie in their relatively big sample size, the homogeneously measured propertied, the depth of the observations and the representativeness of galaxies within the selected stellar mass range \citep{Saintonge2022}. In this study we mainly use the sample of xCOLD GASS, in which 290 galaxies are detected in $\hi$ ($HI\_FLAG=$0, 1 or 2) without confusion ($HIconf\_flag=$0), and 232 galaxies are also detected in CO. 

We select the disk-dominated, non-AGN galaxies from the xCOLD GASS sample. We require the galaxies to have optical r-band concentration $\rnine/\rfive<2.7$. These selections result in a subsample of xCOLD GASS of 172 galaxies. After excluding the AGN with the identification criteria of \citet{Baldwin1981}, \citet{Kewley2001}, and \citet{Kauffmann2003}, we obtain 156 galaxies, which are referred to as the xCOLD GASS selected sample. 

We use the upper limits of gas masses for analysis when galaxies are undetected in gas (either $\hi$ or CO). There are 22 non-detection in CO, 10 non-detection in $\hi$ and 11 galaxies both non-detection in CO and $\hi$. The dust masses of xCOLD GASS galaxies will be estimated using the method developed in Section \ref{subsec:predict mass}.

Spatially consistent estimate of the gas mass is important in the analysis of this paper. In the xCOLD GASS data release, to obtain the total CO flux, \citet{Saintonge2017} applied the prescription of aperture correction presented in \citet{Saintonge2012}, which is based on an exponential 2D model of the molecular gas disk. In this paper, we apply a different aperture correction to our sample, following the method of \citet{Boselli2014} applied to the HRS data. This different correction method considers the 3-D distribution of the molecular gas and extrapolates central single-beam observation of extended galaxies. Compared with the 2D method, the 3D method provides more accurate aperture calibration \citep{Boselli2014}. The median aperture correction across our main sample is 1.25. We use the same conversion factor $\alphaCO$ as in Section \ref{subsec:HRS selection} to calculate $\mHtwo$ based on the CO (1-0) luminosity.

\subsection{JINGLE}\label{subsec:JINGLE}
JINGLE is a survey of 193 galaxies designed to study the relations between cold ISM and global galaxy properties \citep{Saintonge2018}. It combines SCUBA-2 observations at 850 $\mu \rm m$ from the JCMT \citep{Smith2019} with the IR data from $\it Herschel$ to obtain accurate dust mass \citep{Lamperti2019,Looze2020}. About half of the galaxies in the sample have CO(2-1) line observations with the RxA3m instrument of the JCMT, and have $\hi$ data from ALFALFA or a supplementary Arecibo program (PI: M.W.L. Smith) that is not released yet. Similar to the xCOLD GASS sample, these galaxies have $\mstar > 10^{9} \msun$ and $0.01<z<0.05$. Same as xCOLD GASS, we select the disk-like galaxies and exclude the AGNs. The resultant sample has 119 galaxies, and is referred to as the JINGLE selected sample.

The dust masses of JINGLE galaxies are taken from JINGLE IV \citep{Looze2020}, which are estimated through Bayesian fitting of dust SED models. 
In order to compare the HRS-related dust masses with those of JINGLE, we correct for a systematic difference, which increases the dust mass of JINGLE by 0.185 dex. This systematic bias comes from the difference in derived dust masses between the \citet{Draine2007} model used in \citet{Ciesla2014} and THEMIS model \citep{Jones2013,Jones2017} used in \citet{Looze2020} for the HRS sample.

\subsection{Additional data for xCOLD GASS and JINGLE}
We obtain additional multiwavelength measurements for the xCOLD GASS and JINGLE selected samples from public databases. We do not use similar parameters published along with the catalogs of these two surveys to avoid the possible inconsistency between different measurements. 

The optical parameters are retrieved from the SDSS DR7 database \citep{Abazajian2009}. We obtain stellar mass and flux measurements of strong emission lines ($H\alpha$, $H\beta$, $\Nii \lambda$ 6585, $\oii \lambda\lambda$ 3727, 3729 and $\oiii \lambda$ 5007) from the SDSS DR8 MPA-JHU catalog \citep{Tremonti2004, Kauffmann2003}. Dust attenuation for emission lines are corrected by using the Balmer decrement ($H\alpha/H\beta$) and assuming the Calzetti attenuation law \citep{Calzetti2000}. As in HRS, we adopt the O3N2 index as the estimate of gas-phase metallicity. The SFRs are taken from the GSWLC-A2 \citep[GALEX-SDSS-WISE legacy catalog; ][]{Salim2016, Salim2018}, which are estimated through SED fitting. There are 9 galaxies from xCOLD GASS and 11 galaxies from JINGLE which are not included in GSWLC and are thus excluded from the samples. We exclude the galaxies without reliable measurements of metallicity (16 galaxies from xCOLD GASS and 8 galaxies from JINGLE) and stellar surface density (only 1 galaxy from xCOLD GASS). Then, we exclude 5 JINGLE galaxies that deviated more than 1 dex from the SFR-WISE W3 luminosity relation, which may be caused by contamination of the W3 observation or erroneous SFR estimates. After all these exclusions to ensure availability in multi-wavelength measurements, the sample contains 247 galaxies, of which 218 galaxies have directly detected measurements (125 from xCOLD GASS and 93 from JINGLE) in CO, $\hi$ and W3 flux. There are 9 galaxies with CO upper limits and 6 galaxies with $\hi$ upper limits from xCOLD GASS. 

Because the fraction of excluded galaxies from the above selection procedure is relatively small ($<1/5$), the final sample can be viewed as a relatively complete disk-dominated subset from the original xCOLD GASS (JINGLE) selected sample . We refer to it the xCOLD GASS (JINGLE) disk sample.

\begin{deluxetable*}{lccccccc}
\tablecaption{The median values for three disk sample properties.\label{tab:data}}
\centering
\tabcolsep=5pt
\renewcommand\arraystretch{1.2}
\tablehead{
\colhead{Sample$^{a}$} & \colhead{$\rm 12+\log (O/H)$} & \colhead{$\log$\,$M_{\star}$ } & \colhead{$\log$\, SFR} & \colhead{$\log$\,$\mdust$ $^{b}$}& \colhead{$\log$\,$\mhi$ } & \colhead{$\log$\,$\mHtwo$ } & \colhead{$\log$\,$\mgas$}\\
\colhead{} & \colhead{} & \colhead{(M$_{\odot}$)} & \colhead{(M$_{\odot}$ yr$^{-1}$)} & \colhead{(M$_{\odot}$)} & \colhead{(M$_{\odot}$)} & \colhead{(M$_{\odot}$)} & \colhead{(M$_{\odot}$)}
}
\startdata
HRS (55) & $8.63\pm0.02$ & $9.53\pm0.04$ & $-0.14\pm0.08$ & $7.12\pm0.06$ & $9.29\pm0.07$ & $8.77\pm0.05$ & $9.61\pm0.06$\\
xCOLD GASS (125) & $8.73\pm0.01$ & $9.92\pm0.05$ & $-0.09\pm0.07$ & $7.32\pm0.07$ & $9.49\pm0.06$ & $8.85\pm0.06$ & $9.82\pm0.08$\\
JINGLE (93) & $8.75\pm0.01$ & $10.14\pm0.06$ & $0.13\pm0.06$ & $7.53\pm0.04$ & -- & -- & --\\
\enddata
\begin{flushleft}
Notes. $^{a}$ The value in each parenthesis represents the number of galaxies in each of the three disk samples.

$^{b}$ Taking into account the systematic difference caused by the different dust models, we apply a correction to the dust mass of JINGLE. The dust mass of xCOLD GASS is based on our prediction in Section \ref{sec:dust mass}.
\end{flushleft}
\end{deluxetable*}

\subsection{WISE photometry}\label{subsec:WISE photometry}
We measure the WISE W3-band luminosities for the xCOLD GASS and JINGLE disk samples. In order to remove the stellar contributions and apply flux corrections to the W3 band, we perform photometry for WISE W1, W2, and W3 bands together.

The WISE provides a full sky survey in the 3.4, 4.6, 12, and 22 $\mu \rm m$\footnote{\citet{Brown2014} found that the effective wavelength of WISE W4 band should be revised to 22.8 $\mu \rm m$} MIR bands. For each band, the spatial resolution is about $5.9"$, $6.5"$, $7.0"$, and $12.4"$ respectively. The images of WISE are downloaded from the NASA/IPAC Infrared Science Archive\footnote{\url{https://irsa.ipac.caltech.edu/frontpage/}}. We perform aperture photometry using the \verb|Astropy photutils| package of Python \citep{Bradley2019}. The adopted aperture sizes for all the WISE bands are the same and are twice the petrosian radius in the SDSS $r$-band, which should enclose most of the fluxes in galaxies. We also use the upper limits for W3 non-detections by calculating 3-$\sigma$ flux within twice the r-band petrosian radius. There are 3 such W3 undetected galaxies in xCOLD GASS disk sample and 2 in JINGLE disk sample. 

We follow the instructions from Explanatory Supplement to the WISE Preliminary Data Release Products\footnote{\url{http://wise2.ipac.caltech.edu/docs/release/prelim/ expsup/wise_prelrel_toc.html}} to correct for systematic biases due to aperture effects and SED dependent flux calibration offset. 
For the galaxies in our sample which are typically extended, two corrections should be applied: (1) aperture correction: $-$0.034, $-$0.041, and 0.03 in the W1, W2, and W3 band, respectively; (2) color correction: a discrepancy between red and blue sources due to the wide WISE band \citep{Wright2010,Jarrett2011,Jarrett2013}.

In addition to ISM emission, the W3 band also has contamination from the evolved stellar population. We adopt the same factor of 15.8\% for the W1 light to be contained in the W3 as \citet{Cluver2017}. They estimated this value with the method of \citet{Helou2004}. After subtracting the stellar emission and converting flux density to spectral luminosity, we get the luminosity $\LWthree$ for the W3 band. We emphasize that this luminosity is not for the total flux within the band, but rather represents the ISM emission.

\subsection{Estimating $\hi$ Mass within the Optical $\rnine$}\label{subsec:inner HI}
We follow the method in \citet{Wang2020}, which estimates the $\hi$ mass within the optical radius $\rnine$ from the integral $\hi$ mass $\mhi$. First of all, for the given $\mhi$ of a galaxy, we estimate the characteristic radius $\rhi$ based on the relation between $\dhi$ and $\mhi$ \citep{Swaters2002,Wang2016}. Then, we estimate the $\hi$ mass beyond the optical $\rnine$ ($\mhiout$) based on the observational fact that the $\hi$ surface density $\Sigma_{\hi}$ profiles of disk-like galaxies \citep{Wang2014} exhibit similar shapes in the outer region, when the radius is normalized to $\rhi$. 
We use the median $\hi$ surface density profile from \citet{Wang2016}, which was derived from a sample of 168 nearby spiral and dwarf galaxies. The maximum radius of original median profile has been extrapolated from 1.15 $\rhi$ out to 1.5 $\rhi$. If $\rnine > 1.5\rhi$, it means that the $\hi$ gas is concentrated in the optical disk of galaxy, and $\mhiout = 0$. After subtracting $\mhiout$ from $\mhi$, we obtain the $\hi$ mass within $\rnine$, $\mhiin$. To ensure that galaxies have similar $\hi$ surface density profile shapes in the outer disks, it's important to select the disk dominated galaxies by requiring the $r$-band light concentration $\rnine/\rfive <2.7$ \citep{Wang2020}, as we have done in Section \ref{subsec:HRS selection}. 

Then we derive the inner gas mass: $M_{\rm gas,in} = 1.36\times(M_{\rm \hi,in}+\mHtwo)$. The parameter 1.36 accounts for the contribution from helium. The uncertainties of $\mgasin$ are propagated from $\mhiin$ and $\mHtwo$. 

\section{The relations between gas-to-dust ratio and metallicity}\label{sec:gas-to-dust ratio}
We use the HRS disk sample to study the dependence of gas-to-dust ratio on the gas phase metallicty. The mass ratio of gas in different phases ($\hi$, $\htwo$, and total gas) over dust mass is generally considered to vary with gas-phase metallicity \citep{Leroy2011,Remy-Ruyer2014,Berta2016,Janowiecki2018}. The trend of integral gas-to-dust mass ratio with metallicity is often used to constrain physical processes governing galaxy evolution \citep{Draine2007,Remy-Ruyer2014,Galliano2018}. 

The data set analyzed involve upper limits, so we use the survival analysis method to deal with such censored data \citep{Feigelson1985,Isobe1986}. Such an analysis methodology is used throughout this paper unless otherwise pointed out. 
For correlation based analysis, we mainly use the Kendall's $\tau$ coefficient ($R_{k}$) to quantify the rank correlation between two sets of (possibly censored) data, while the Pearson's correlation coefficient ($R_{p}$) is also used as a reference to quantify the linear correlation without considering upper limits. 
The $R_{k}$ is computed using the \verb|cenreg| routine in the \verb|NADA R| package\footnote{\url{https://CRAN.R-project.org/package=NADA}}. 
To obtain linear relations, we perform Akritas–Theil–Sen (ATS) regression to the censored data. It was proven that ATS estimator can give reliable linear fits, as long as the censored part does not dominate the sample \citep[e.g.,][]{Stark2021}. 
We use bootstrapping method to estimate the uncertainties of the slopes, intercepts, $R_{p}$, and $R_{k}$. In practice, we resample allowing repetition with the sample size fixed, and derive the statistical parameter in question with the new sample; we repeat this process 1000 times and obtain 1000 sets of the parameter; the confidence intervals are derived from 16 and 84 percentiles of this distribution for that parameter.

\subsection{Gas-to-dust ratio for inner $\hi$ and inner gas}\label{subsec:gtd ratio}
\begin{figure}[htbp]
	\centering
	\includegraphics[width=1\linewidth]{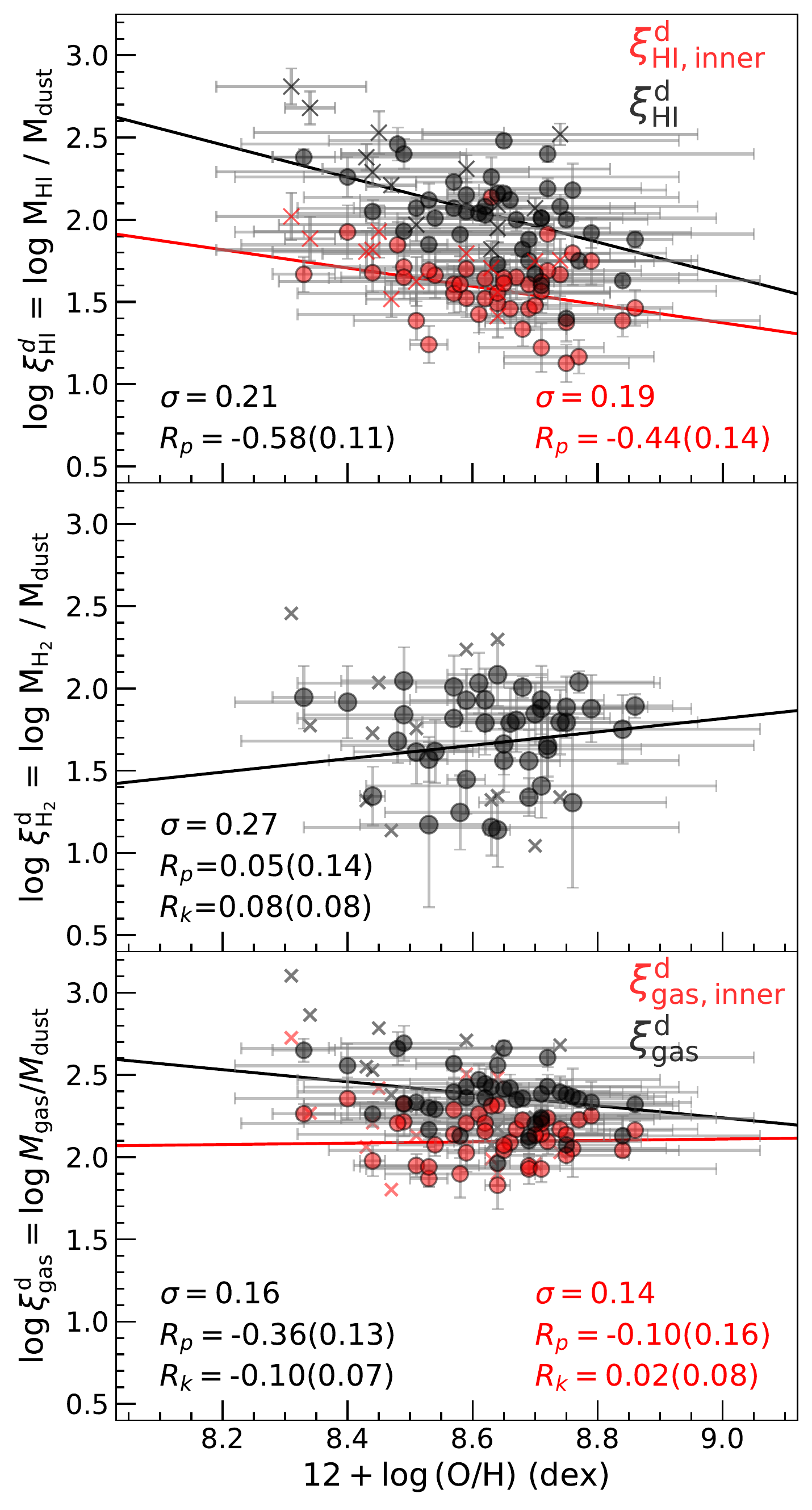}
	\caption{Gas-to-dust ratio as a function of the gas-phase metallicity with (red) or without (black) outer $\hi$ gas removed. Top: $\xihi$ and $\xihiin$ ; middle: $\xihtwo$; bottom: $\xigas$ and $\xigasin$. In each panel, best-fit linear lines, Pearson's correlation coefficients ($R_{p}$, and their uncertainties estimated from bootstrapping), Kendall's $\tau$ coefficients ($R_{k}$, and their uncertainties), and 1-$\sigma$ scatters are shown in red and black, respectively.}
	\label{fig:gas-to-dust_ratio.pdf}
\end{figure}

The gas-to-dust ratios of integral $\hi$ ($\xihi$), $\htwo$ ($\xihtwo$), and integral gas ($\xigas$) of galaxies as a function of gas-phase metallicity are presented as black symbols in Figure \ref{fig:gas-to-dust_ratio.pdf}. The crosses denote the galaxies undetected in the CO line. Noted that these galaxies are detected in the $\hi$, so the crosses in the upper panel do not represent upper limits. 

The black solid lines represent the best-fit linear relations. The scatters, $R_{p}$, and $R_{k}$ are shown in the lower left corner in black color. The trends of $\xihi$ and $\xigas$ are consistent with the relations from \citet{Janowiecki2018}, although we apply slightly more selection criteria on our sample. The $\xihtwo$ show a poorer relation with metallicity than that in \citet{Janowiecki2018}, and the difference mainly comes from the different conversion factor $\alphaCO$ adopted. The metallicity-dependent $\alphaCO$ in this work tends to result in smaller and higher $\mHtwo$ for metal-rich and metal-poor galaxies, respectively, compared to the values in \citet{Janowiecki2018}. Note that the relationship of $\xigas$ is the tightest one, whereas that of $\xihtwo$ has the largest scatter.

One source of uncertainty that could bias this relation is that the existence of extended $\hi$ may overestimate the gas-to-dust ratio \citep{Remy-Ruyer2014,Casasola2020}. Then, using our estimated inner $\hi$ mass $\mhiin$ and inner gas mass $\mgasin$ (we presume that the molecular gas mainly resides at inner region), we minimize the effect of outlying $\hi$ mass on the gas-to-dust ratio. The inner gas-to-dust ratio ($\xihiin$ and $\xigasin$) relations and best-fit lines are shown in red colors in Figure \ref{fig:gas-to-dust_ratio.pdf}. The $R_{p}$, $R_{k}$, and 1-$\sigma$ scatter are also given in red color in this figure.

In the top panel of Figure \ref{fig:gas-to-dust_ratio.pdf}, we can see that the trend of $\xihiin$ is slightly weaker compared to that of $\xihi$. The relationship of $\xihiin$ shows a smaller scatter of 0.19 dex compared to 0.21 dex of $\xihi$. 

More importantly, from the bottom panel of Figure \ref{fig:gas-to-dust_ratio.pdf}, we can see that $\xigasin$ show different trends from $\xigas$ with metallicity. The $\xigas$ show an anti-correlation with $\rm 12+\log (O/H)$, but $\xigasin$ do not significantly depend on $\rm 12+\log (O/H)$ ($R_{p}=-0.10$ and $R_{k}=0.02$). Furthermore, among all the relations, the scatter in the best-fit relation between $\xigasin$ and $\rm 12+\log (O/H)$ is the smallest (0.14 dex). The best-fit relation of $\xigasin$ is given as the following equation:

\begin{eqnarray}
\log \xigasin = \log \frac{\mgasin}{\mdust} = && (1.73\pm2.01) + (0.04\pm0.23) \nonumber \\
&& \times (12+\log \text{(O/H)}).
\end{eqnarray}

We recall that the anti-correlation between $\xigas$ and metallicity might come from the dilution of outer $\hi$, since we presume that the outer region is beyond the region where the dust mass is detected. Now when we remove the outer $\hi$, the negative correlation is weakened or disappear. 
Such an almost metal-independent gas-to-dust ratio indicates promising application in deriving inner gas mass from the dust mass, and vice versa. The inner gas mass has a unique but overlooked role in studying galactic star formation and galaxy evolution. The formation of $\htwo$ molecules from $\hi$ has been found to be inefficient whereas interstellar dust grains can act as a catalyst to increase the efficiency. In the outer region of galaxies, the more extended $\hi$ distribution is hardly responsible for the formation of $\htwo$ and stars \citep{Bertemes2018,Wang2020}. Therefore, it is the inner gas which provides the direct reservoir of material for star formation. 

It is worth mentioning that \citet{Casasola2020} with the DustPedia sample also derived a relation of inner gas-to-dust mass ratio as a function of metallicity. A major difference here is that, we use $\rnine$, while they use $r_{25}$ as the division for inner disks (the mean value of $\rnine/r_{25}$ is about 0.70 in our sample), and we derive $\mhiin$ in a different way. We use the median $\hi$ profile of 168 galaxies from \citet{Wang2016} and we derive $\mhiin$ by subtracting the integrated $\hi$ mass in the outer region where different galaxies have similar profile shapes when the radius is normalized by $\rhi$, whereas \citet{Casasola2020} used the $\hi$ surface density model obtained from the average of 42 galaxies in \citet{Wang2014} and directly integrated the model out to optical radius. In addition, there are differences in the dust models assumed to derive $\mdust$, metal-dependent $\alphaCO$ used to derive $\mHtwo$, and calibrations of gas-phase metallicity, between these two studies. Possibly because of these differences, the relation presented in \citet{Casasola2020} show much steeper slopes.

The result is also in agreement with \citet{Bertemes2018}. They find that the relation between CO and dust emission can be much tighter by introducing in a metal-dependent correction, and suggest that such a correction accounts for $\hi$ gas in the same region. 

\subsubsection{Gas-to-dust mass ratios with different radius to derive the inner $\hi$}
Due to the inhomogeneous extents of gas and dust, the measurement of the gas-to-dust ratio depends on the aperture size. In our sample, the 85th, 50th, and 15th percentiles of the $\rhi/\rnine$ are 3.429, 2.299, and 1.693, respectively. So the choice of radius used in the method to estimate inner $\hi$ mass has an effect on the relationship between gas-to-dust ratio and metallicity. 

\begin{figure}[htbp]
	\centering
	\includegraphics[width=1\linewidth]{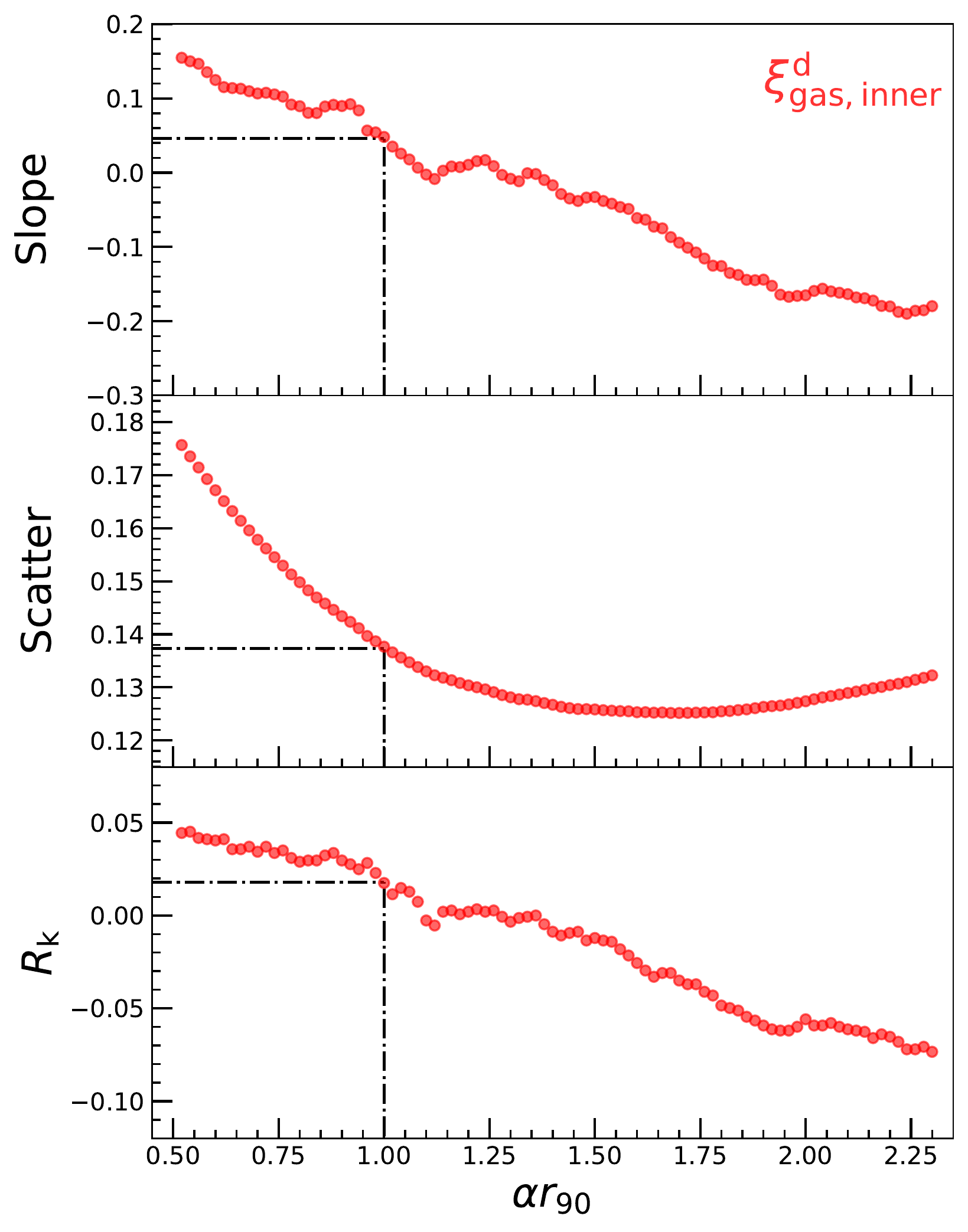}
	\caption{Effects of different cutting radii $\alpha \rnine$ on the relations between inner gas-to-dust ratio and metallicity. Top: slope; middle: scatter; bottom: Kendall's $\tau$ coefficients. The dashed black lines indicate the positions where $\alpha = 1.0$.}
	\label{fig:alpha_relation.pdf}
\end{figure}

Therefore, we test different dividing radius $\alpha \rnine$ for the inner region, with $\alpha$ ranging from 0.5 to 2.3, and study the variation of slopes, scatters, and the Kendall's $\tau$ coefficients of the relations between inner gas and metallicity studied in Section \ref{subsec:gtd ratio}. The results are shown in Figure \ref{fig:alpha_relation.pdf}. In each panel, dashed lines show the positions where $\alpha = 1.0$. Both the $R_{k}$ and slope change monotonically from positive to negative values as $\alpha$ increases. The scatter reaches its minimum around 1.6 $\rnine$.

The trends show that 1.0 $\rnine$ is a good option to derive a relatively metal-independent inner gas-to-dust ratio. Considering measurement uncertainties, it is very close to the radius (around 1.1 $\rnine$) where the slope and $R_{k}$ of the related metallicity--$\xigasin$ relation are close to 0. Meanwhile, the scatter of the related relation is not significantly different (only by about 0.01 dex) from that of minimum value at around 1.6 $\rnine$. 

\section{Predicting inner gas mass with dust mass and infrared luminosity}\label{sec:relation}
 In this section, we derive direct relations between the inner gas mass and dust mass, as $\xigasin$ depends only weakly on metallicity. 
The relations can be used as predictor of $\mgasin$ based on $\mdust$, and vice versa. 

It is also useful to derive and examine the relation between gas and IR luminosity since the latter is easier to obtain than $\mdust$. Previous studies have established the tight relationship between far infrared (FIR) luminosity and $\mgas$ \citep[e.g.,][]{Scoville2014,Groves2015,Scoville2016}. Among FIR luminosities, the $\Lfive$ shows the strongest correlation based on the results of \citet{Janowiecki2018}. So here we focus on $\Lfive$ as the FIR luminosity.

We also compare these relations to those of integral gas masses to highlight the advantage of using inner gas masses.

\begin{figure*}[htbp]
	\centering
	\includegraphics[width=1\linewidth]{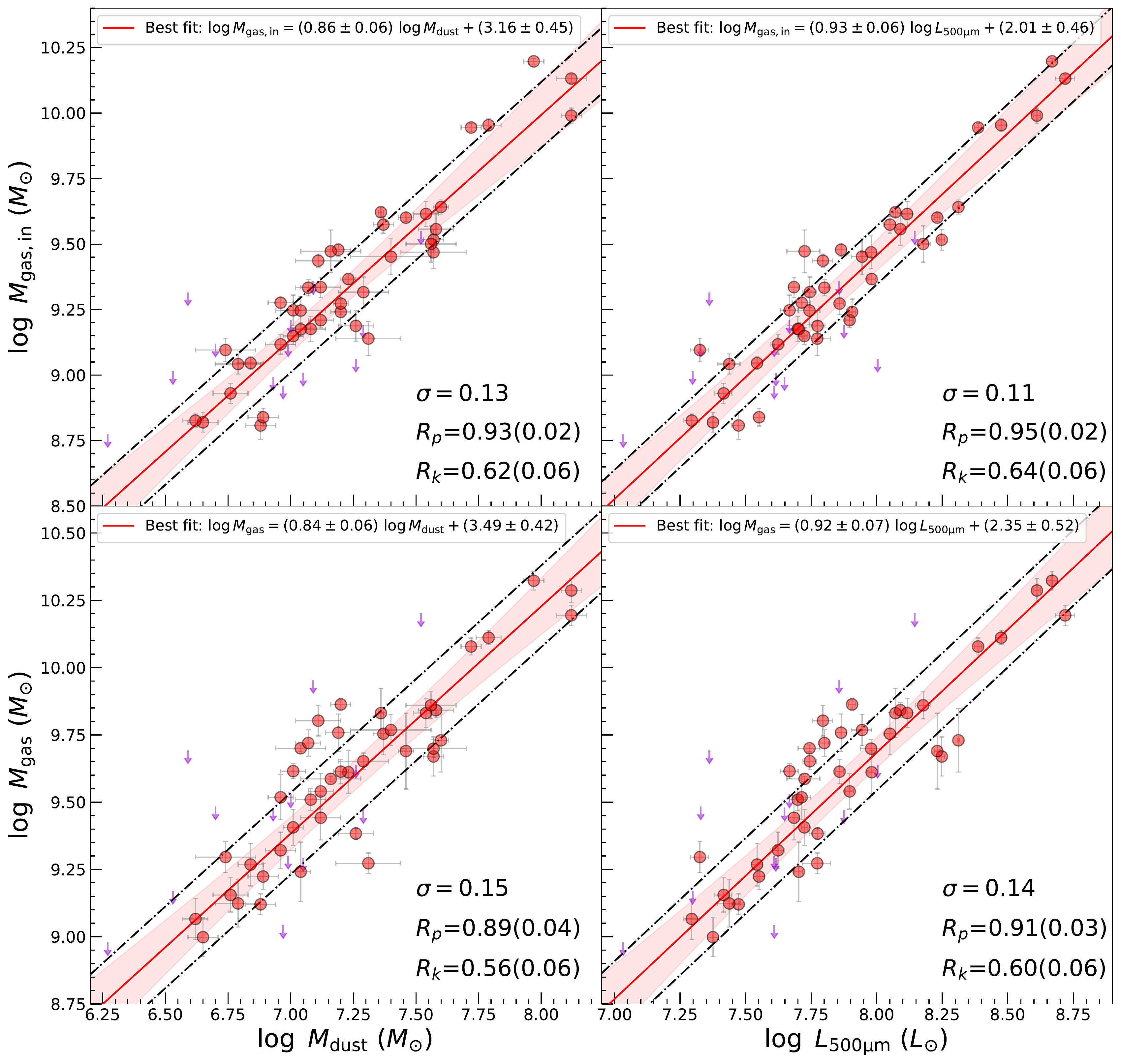}
	\caption{The relations between gas parameters (gas mass and inner gas mass) and dust parameters (dust mass and FIR 500 $\mu m$ luminosity). The red points denote galaxies with CO detections while the purple arrows denote the 3-$\sigma$ upper limits. The gray error bars show the measurement uncertainties of each galaxy. The red lines and two black dashed lines show, respectively, the ATS regression line to the data and the 1-$\sigma$ scatter from the fitting lines. The red shaded region represents the confidential band of the regression line.} The scatters are computed for CO detected galaxies and also shown in the lower right corner for each panel. The Pearson's correlation coefficients ($R_{p}$), Kendall's $\tau$ coefficients ($R_{k}$) and their uncertainties for the whole sample are also displayed in the lower right corner.
	\label{fig:innergas_dust.pdf}
\end{figure*}

\subsection{Predicting inner gas mass with $\mdust$}\label{subsec:predict mass}
The left column of Figure \ref{fig:innergas_dust.pdf} shows the relations between the masses of gas ($\mgasin$ and $\mgas$) and dust ($\mdust$). The scatters, Pearson's correlation coefficients, and Kendall's $\tau$ coefficients are listed in the right corner of each panel. The median values and uncertainties from bootstrapping are listed in Table \ref{tab:Relation}. 

The slopes of both relations are sub-linear (0.86 and 0.84). 
The relation of $\mgasin$ with $\mdust$, as expected, is tighter (with higher correlation coefficients, and the linear fit with smaller scatter) than that of $\mgas$. The best-fit relation between $\mgasin$ and $\mdust$ is given below:
\begin{eqnarray}\label{equation:inner gas-dust}
\log (\mgasin&&/\msun) =   (3.16\pm 0.46)\nonumber \\
&& +(0.86\pm 0.06) \times \log (\mdust/\msun).
\end{eqnarray}

Reversely, to predict $\mdust$ from $\mgasin$, we fit the following equation: 
\begin{eqnarray}\label{equation:dust-inner gas}
\log (\mdust&&/\msun) =  -(2.27 \pm 0.79)\nonumber \\
&& +(1.01 \pm 0.09) \times \log (\mgasin/\msun) ,
\end{eqnarray}
with a 1-$\sigma$ scatter of 0.14 dex.

We also compare the correlations of observed gas masses with gas masses derived using different gas-to-dust ratios: metal-dependent ratio $\xi_{\rm gas,OH}$ and constant ratio $\xi_{\rm gas,c}$. The statistical results are listed in Table \ref{tab:Relation}. The gas masses derived with $\xi_{\rm gas,OH}$ ($\xi_{\rm gas,OH} \times \mdust$) show slightly tighter ($\sigma=0.14$ dex) and stronger ($R_{p}=0.90$ and $R_{k}=0.58$) correlation with the observed $\mgas$ than the gas masses derived with $\xi_{\rm gas,c}$ ($\xi_{\rm gas,c} \times \mdust$), suggesting that a metallicity-dependent gas-to-dust ratio is preferred when deriving $\mgas$ based on $\mdust$. However, this is not the case for $\mgasin$ as the associated $\xigasin$ depends very weakly on the metallicity.

The dust mass has been found to be tightly correlated with gas mass on both global \citep[e.g.,][]{Corbelli2012,Janowiecki2018,Casasola2020} and spatially resolved scales \citep[e.g.,][]{Foyle2012,Abdurro'uf2022b,Casasola2022}. Our results demonstrate that inner gas shows stronger relations than the integral gas mass with dust mass.

\subsection{Predicting inner gas mass with $\Lfive$}
In the right column of Figure \ref{fig:innergas_dust.pdf}, we show the relation of gas mass ($\mgasin$ and $\mgas$) versus $\Lfive$. We find that both $\mgasin$ and $\mgas$ are tightly correlated with $\Lfive$. Compared to $\mdust$, $\Lfive$ seems to be an even better estimator for $\mgas$, which was found by \citet{Janowiecki2018}. Here the same is found for $\mgasin$, which is better predicted by $\Lfive$ than by $\mdust$.
Again, $\mgasin$ shows smaller scatters and larger correlation coefficients ($\sigma=0.11$ dex, $R_{p} = 0.95$, and $R_{k} = 0.64$) than $\mgas$ ($\sigma=0.14$ dex, $R_{p} = 0.91$, and $R_{k} = 0.60$). The best-fit relation between $\mgasin$ and $\Lfive$ is given below:
\begin{eqnarray}\label{equation:inner gas-L500}
\log (\mgasin&&/\msun) = (2.01 \pm 0.46)\nonumber \\
&& +(0.93\pm0.06) \times \log (\Lfive/\lsun) .
\end{eqnarray}

We also consider correlations of gas masses with 100, 160, 250 and 350 $\mu m$ luminosity, all of which are worse than the correlations with $\Lfive$ (both for $\mgas$ and $\mgasin$). 
We confirm the results in the literature \citep{Corbelli2012,Groves2015,Janowiecki2018} that $\mHtwo$ shows stronger correlation with $\Lsixteen$ and $\Ltwofive$, while $\mhi$ shows stronger correlation with $\Lfive$. But we additionally find that $\mgasin$ shows stronger correlation with $\Lfive$ than $\mHtwo$. \citet{Janowiecki2018} attributed the similar trends of integral gas and $\hi$ to the fact that $\hi$ is the dominant component in their sample. In our study, after removing the $\hi$ in the outer region, the $\mhiin$ is slightly smaller than $\mHtwo$ ($\log \mhiin = 8.82$ and $\log \mHtwo = 8.84$ in median), but the correlation of $\mhiin$ with $\Lfive$ is still the strongest among its correlations with FIR luminosities. These results demonstrate that FIR luminosities, especially $\Lfive$, display close connections with total gas (i.e. not just $\hi$) in the inner region, which is consistent with spatially resolved relations \citep[e.g.,][]{Foyle2012,Abdurro'uf2022b,Casasola2022}.

\begin{figure*}[htbp]
	\centering
	\includegraphics[width=\linewidth]{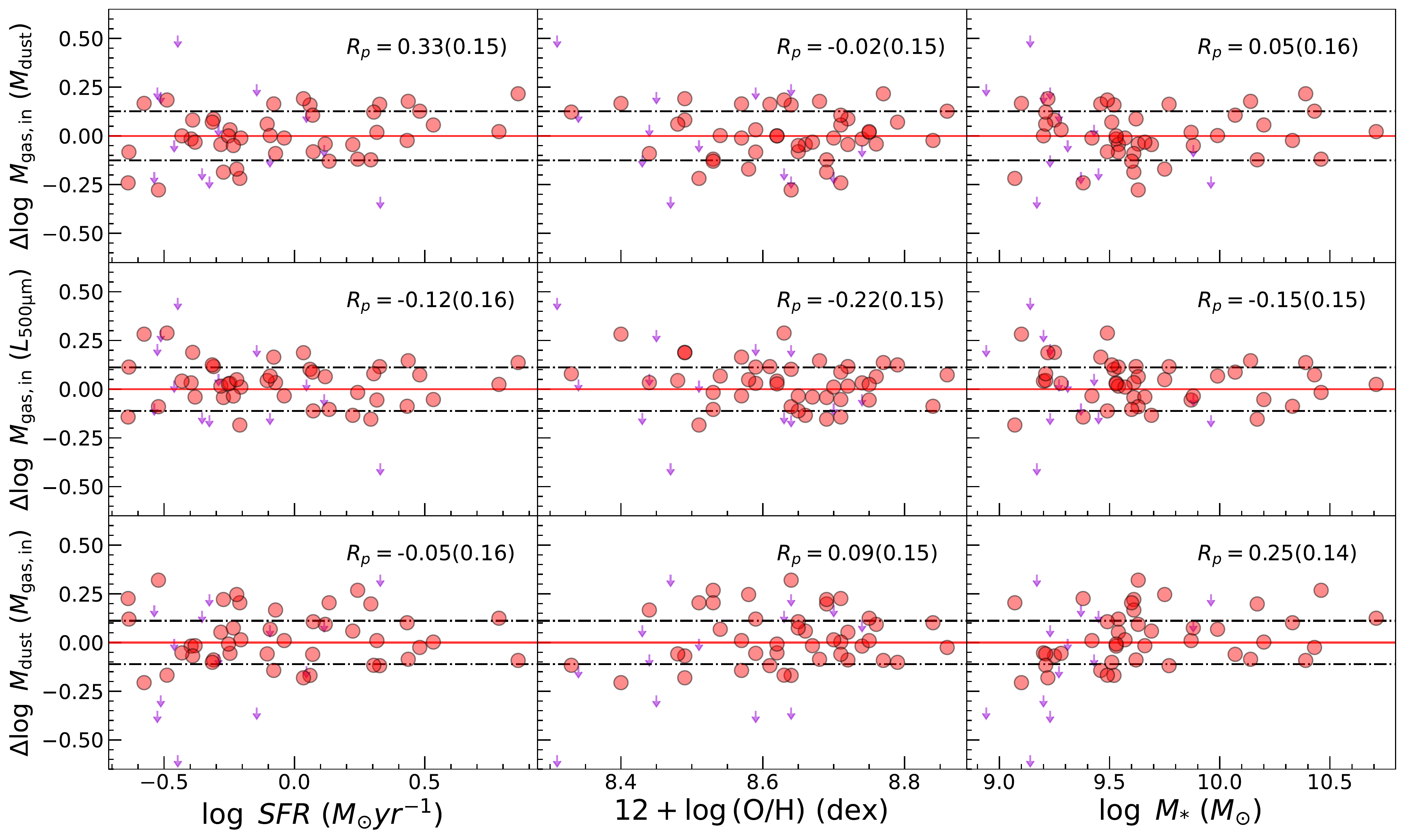}
	\caption{Offsets of inner gas masses or dust masses from the values predicted with the $\mdust$--$\mgasin$ relations. The top row plots the offset of inner gas masses from values predicted with dust masses ($\Delta \log(M_{\rm{gas,in}})(M_{\rm{dust}})$), the middle row the offset of inner gas masses from values predicted with $\Lfive$ ($\Delta \log(M_{\rm{gas,in}})(\Lfive)$), and the bottom row the offset of dust masses from values predicted with inner gas masses ($\Delta \log(M_{\rm{dust}})(\mgasin)$). These offsets are plotted as a function of SFR, gas-phase metallicity and stellar mass. The red circles are CO detections and the purple arrows show 3-$\sigma$ upper limits. The red and two dashed black horizontal lines show the position of 0 and 1-$\sigma$ scatter measured in Figure \ref{fig:innergas_dust.pdf}. The Pearson's correlation coefficient and its uncertainty from bootstrapping are shown in the top right corner of each panel.}
	\label{fig:residual.pdf}
\end{figure*}

\begin{deluxetable*}{lccccc}
\tablecaption{The statistical values for the relations between $\mgasin$ and $\mdust$, $\mgas$ and $\mdust$, $\mgas$ and $\xigas \times \mdust$, $\mgasin$ and $\Lfive$, and, $\mgas$ and $\Lfive$.\label{tab:Relation}}
\centering
\tabcolsep=5pt
\renewcommand\arraystretch{1.2}
\tablehead{
\colhead{} & \colhead{$\mgasin$-$\mdust$} & \colhead{$\mgas$-$\mdust$}  &  \colhead{$\mgas$-$\xi_{\rm gas,OH} \times \mdust$} & \colhead{$\mgasin$-$\Lfive$} & \colhead{$\mgas$-$\Lfive$}\\
}
\startdata
Slope & $0.86\pm0.06$ & $0.84\pm0.06$ & $0.87\pm0.06$ & $0.93\pm0.06$ & $0.92\pm0.06$\\
Intercept & $3.16\pm0.46$ & $3.49\pm0.42$ & $1.24\pm0.54$ & $2.01\pm0.46$ & $2.35\pm0.49$\\
Scatter ($\sigma$) & 0.13 & 0.15 & 0.14 & 0.11 & 0.14\\
Pearson's correlation coefficient ($R_{p}$) & $0.93\pm0.02$ & $0.89\pm0.04$ & $0.90\pm0.04$ & $0.95\pm0.02$ & $0.91\pm0.03$\\
Kendall's $\tau$ coefficients ($R_{k}$) & $0.62\pm0.06$ & $0.56\pm0.06$ & $0.58\pm0.06$ & $0.64\pm0.06$ & $0.60\pm0.06$
\enddata
\end{deluxetable*}

\subsection{Residual trends in relations}\label{subsec:residual}
We then examine the residuals which are defined as the offset of dust masses or inner gas masses from the predicted values. Firstly, we conduct a comprehensive series of statistical tests on the residuals, and the results showed that the residuals remain invariant with the independent variables and conform to a Gaussian distribution. These results provide compelling evidence for the appropriateness of our fitting, allowing us to proceed with further analysis of the residuals.
Figure \ref{fig:residual.pdf} shows the residuals of relations obtained in the previous section as a function of SFR, metallicity, and stellar mass. 
The top row shows the residuals of $\mgasin$ related to $\mdust$, the middle row the residuals of $\mgasin$ related to $\Lfive$, and the bottom row the residuals of $\mdust$ related to $\mgasin$.

The residuals of $\mdust$-predicted $\mgasin$ do not have a significant dependence on $\rm 12 + \log(O/H)$ and $M_{*}$, but show a weak correlation with SFR. While defining inner disks with $\rnine$ removes the dependence on metallicity (as shown in Figure \ref{fig:gas-to-dust_ratio.pdf}), the close relationship between inner gas and SFR might still contribute to the scatters of the $\mdust$--$\mgasin$ relation. 

On the other hand, the residuals of the $\Lfive$-predicted $\mgasin$ show nearly no correlation with SFR (within 1-$\sigma$ uncertainty), which may be attributed to the closer connection between FIR luminosities and SFR compared to that of $\mdust$. These residuals exhibit a weak anti-correlation with $\rm 12 + \log(O/H)$, which might hint at a possible secondary correlation with the metal content of ISM for the $\Lfive$--$\mdust$ relation.

For the residuals of $\mgasin$-predicted $\mdust$, there is no correlation with SFR or metallicity, but a weak correlation with $M_{*}$. 

These secondary dependence should be kept in mind when the predicted quantities are used. We conduct linear fits to the three relatively strong trends discussed above, and report the slopes here, which can be used to derive related systematic uncertainties. The slopes for the relations of $\Delta \log M_{\rm{gas,in}}(\mdust)$ versus $\log$ SFR, $\Delta \log M_{\rm{gas,in}}(\Lfive)$ versus $\rm 12+\log(O/H)$, and $\Delta \log \mdust (\mgasin)$ versus $\log M_{*}$ are 0.11, $-$0.21, and 0.09, respectively. We warn about the limited number of galaxies in our sample, and possible dependencies due to statistical fluctuation, which should be tested with more data when available.

On the whole, Equation \ref{equation:inner gas-dust} and \ref{equation:inner gas-L500} provide reasonable predictor of the inner gas mass based on the dust mass and $\Lfive$, and Equation \ref{equation:dust-inner gas} reasonable predictor of the dust mass based on the inner gas mass, for disk-like galaxies.

\section{Predicting dust masses for xCOLD GASS disk sample}\label{sec:dust mass}
In this section, we apply the inner gas--dust mass relation to estimate dust masses for the xCOLD GASS disk sample. We briefly study the relations of the predicted dust mass with other galactic properties. We compare the relations with those in the literature, to justify the robustness of the predicted dust mass. We also compare to the relations of the JINGLE disk sample, to show the less biased nature of the xCOLD GASS disk sample.

We have applied the aperture correction methodology of \citet{Boselli2014} to the xCOLD GASS CO fluxes, and derive $\mHtwo$ in a consistent way as for the HRS disk sample (Section \ref{subsec:xCOLD GASS}).
We have derived the inner $\hi$ mass within the optical $\rnine$ (see Section \ref{subsec:inner HI}) to obtain the inner gas mass. Then we apply our newly calibrated relation between dust mass and inner gas mass (Equation \ref{equation:dust-inner gas}) to predict dust mass for the xCOLD GASS disk sample.

\subsection{Sample properties and dust scaling relations}
In order to get prepared for the comparison of scaling relations, we compare several characteristics properties of xCOLD GASS disk sample with those of JINGLE disk sample. 

\begin{figure*}[htbp]
	\includegraphics[width=\linewidth]{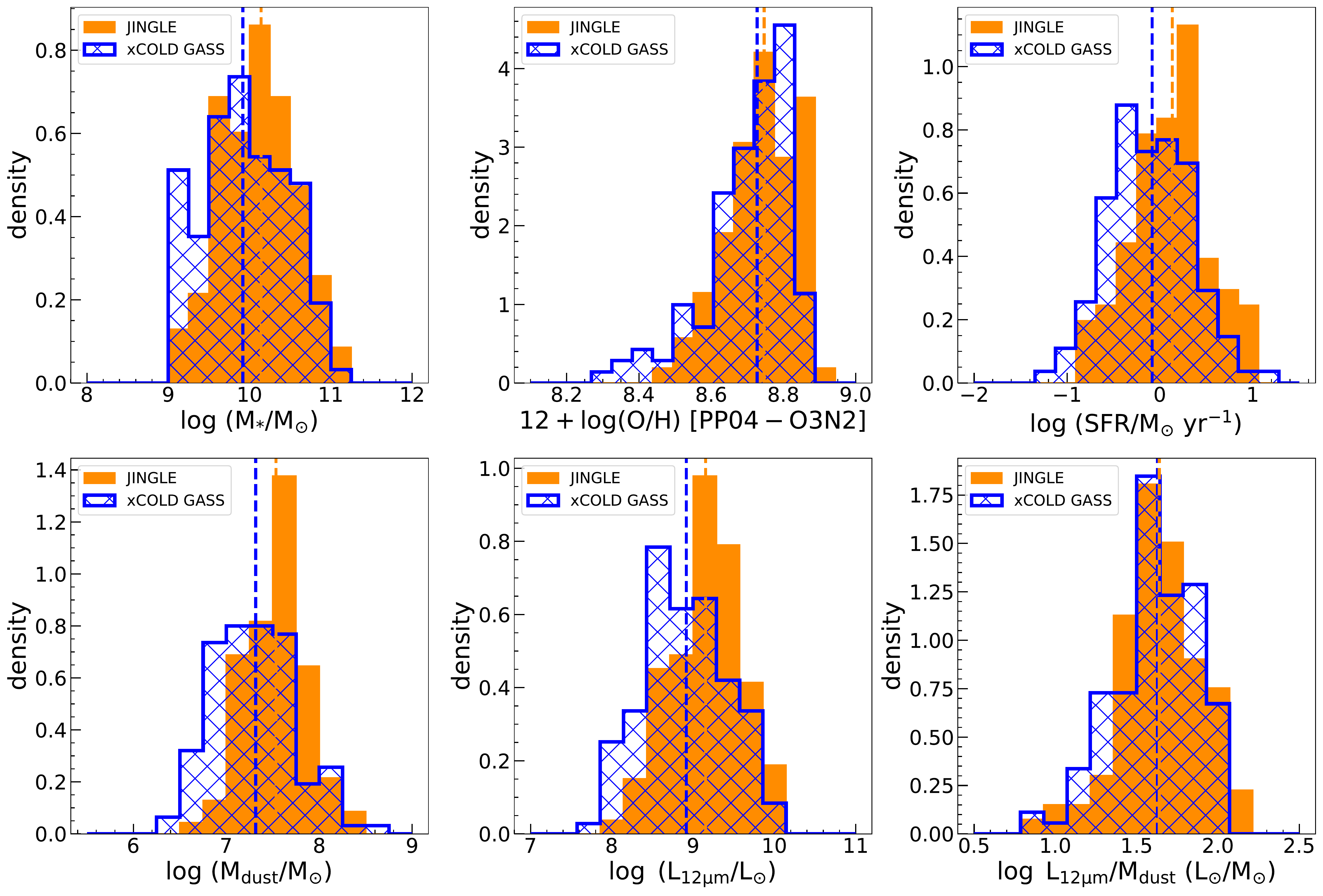}
	\caption{JINGLE and xCOLD GASS disk sample properties. Top row: histograms of stellar masses, gas phase metallicities and SFRs for JINGLE (orange) and xCOLD GASS (blue). Bottom row: histograms of dust masses, WISE 12 $\mu m$ luminosities and $\LWthreeMdust$. The median values are indicated with vertical dashed lines.}
	\label{fig:distribution.pdf}
\end{figure*}

\begin{figure*}[htbp]
	\includegraphics[width=\linewidth]{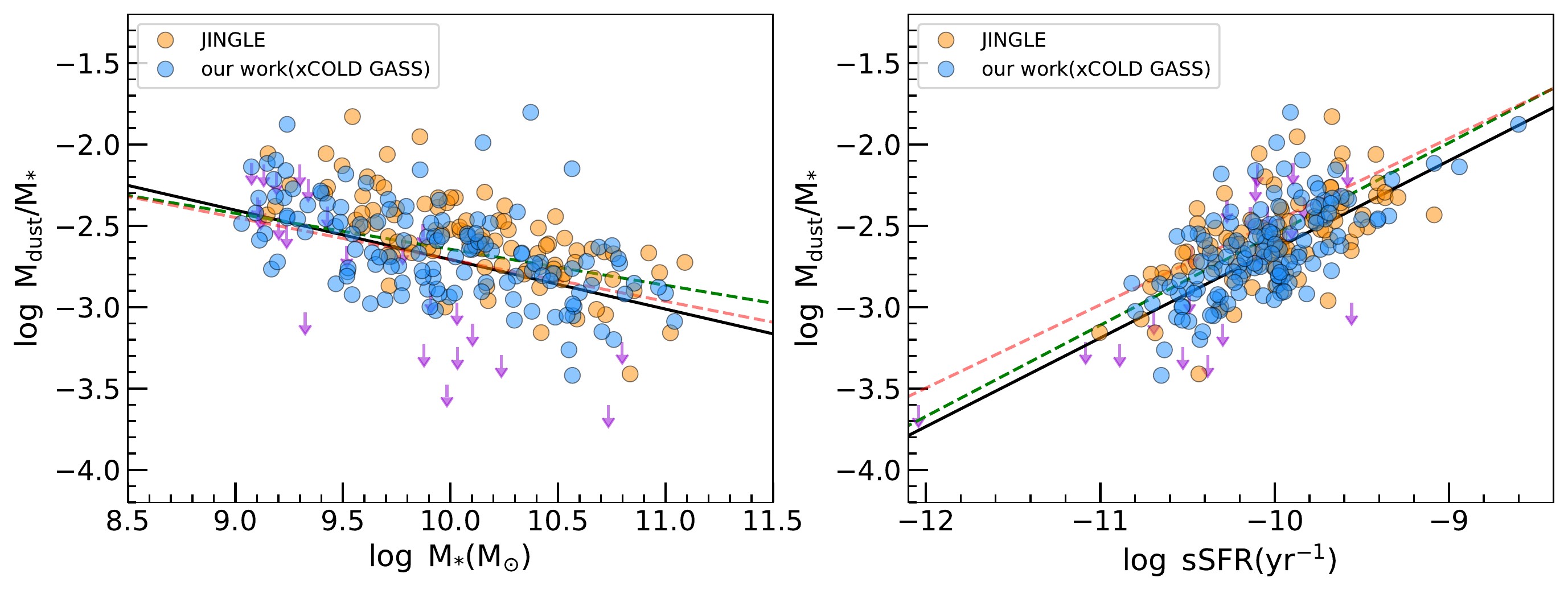}
	\caption{The dust scaling relations for JINGLE (orange) and xCOLD GASS disk sample (blue). Left: the relation between the dust-to-stellar mass ratio ($M_{\rm{dust}}/M_{*}$) and stellar mass. Right: the relation between $M_{\rm{dust}}/M_{*}$ and sSFR. The purple arrows show the dust mass of xCOLD GASS disk sample estimated from 3-$\sigma$ upper limits of CO. The black lines are the best-fit lines only for the xCOLD GASS disk sample and overlap with the scaling relations from \citet{Looze2020} (green dashed line) and \citet{DeVis2017} (red dashed line).}\label{fig:dust_scaling_relation.pdf}
\end{figure*}

Figure \ref{fig:distribution.pdf} shows the distributions of several galactic properties of the xCOLD GASS disk sample and JINGLE disk sample. On the top row, we show the histograms of gas-phase metallicity ($\rm 12 + \log(O/H)$), stellar mass ($\mstar$) and SFR for the two samples. Due to the similar sample selection, both the JINGLE and xCOLD GASS disk samples have a relatively flat stellar mass distribution, but xCOLD GASS disk sample has more low-stellar mass galaxies, resulting in the median value of stellar mass distribution lower by 0.22 dex. Consequently, the xCOLD GASS disk sample also has slightly lower metallicity and lower SFR, by 0.02 and 0.21 dex respectively. The bottom row of Figure \ref{fig:distribution.pdf} shows the distributions of $\mdust$, 12 $\mu m$ luminosity ($\LWthree$), and the ratios of them, $\LWthreeMdust$. The dust mass and 12 $\mu m$ luminosity of JINGLE galaxies are clearly higher than those of xCOLD GASS galaxies, possibly caused by both the systematically higher stellar mass and the selection in FIR bands for the JINGLE sample. Nevertheless, the median $\LWthreeMdust$ of JINGLE ($\log \LWthreeMdust=1.64$) is close to the median value of xCOLD GASS ($\log \LWthreeMdust=1.63$).

Scaling relations of specific dust mass ($\MdustMstar$) provide important observational depiction of galaxy evolution. In Figure \ref{fig:dust_scaling_relation.pdf}, we show the relations of $\MdustMstar$ with stellar masses and specific SFR (sSFR), which have been extensively studied in the literature \citep[e.g.,][]{Cortese2012b,Clemens2013,Clark2015,Calura2017,DeVis2017,Orellana2017,Casasola2020,Looze2020}. The anti-correlation with the stellar mass can be interpreted in terms of the gas-dust-star cycle and indicates that the processes of dust destruction are more efficient than production with galaxy evolution. 
Specially, the relation with sSFR has been found to be insensitive to selections in galaxy morphology or dust richness, indicating a close link between dust and star formation \citep[e.g.,][]{daCunha2010,Looze2020}.

\citet{DeVis2017} and \citet{Looze2020} provided relatively unbiased scaling relations of $\MdustMstar$ with the $\mstar$ and sSFR, by carefully including nearby galaxies with low $\MdustMstar$. We plot their relations in Figure \ref{fig:dust_scaling_relation.pdf} as the reference.
We compare the distribution of xCOLD GASS disk galaxies and JINGLE disk galaxies around these relations in Figure \ref{fig:dust_scaling_relation.pdf}. 
The distribution of xCOLD GASS disk galaxies is relatively uniform around the relations of \citet{DeVis2017} and \citet{Looze2020}, and the best-fit line (black line) is also in close agreement with their relations. In contrast, the JINGE disk galaxies are slightly biased toward high $\MdustMstar$ in the relation with $\mstar$, particularly at the low-$\mstar$ end, as pointed out already in \citet{Looze2020}. The JINGLE disk galaxies distribute similarly as the xCOLD GASS disk galaxies around the relation of $\MdustMstar$ versus sSFR, as $\MdustMstar$ is more fundamentally correlated with sSFR than with $\mstar$.

In summary, the xCOLD GASS disk sample shows consistent dust scaling relations with fiducial relations in the literature. It supports that the $\mgasin$--$\mdust$ relation provides a reasonable prediction of dust mass based on the inner gas mass. Due to the different selections, the xCOLD GASS disk sample may be less biased than the JINGLE disk sample to study dust properties. The xCOLD GASS disk sample, with a relatively simple selection, is thus newly equipped with the predicted $\mdust$ for future study. 

\subsection{W3 as an SFR indicator: relations between $\LWthreeMdust$ and other galactic properties}\label{sec:L12/Mdust}

In this section we use the newly predicted $\mdust$ of xCOLD GASS disk sample, to evaluate the goodness of WISE 12 $\mu m$ luminosity as an SFR indicator. 

The MIR 12 $\mu m$ luminosity is often calibrated to indicate SFR, yet its robustness has been questioned. A major reason that the 12 $\mu m$ luminosity can be used to indicate SFR is because the PAH emissions in the WISE W3 band trace the photo-dissociation region around young stars. However, detailed modeling that extracts the PAH emissions from mapping-mode spectroscopy of well-resolved nearby galaxies suggests that, the relation between PAH luminosity and SFR is possibly influenced by the metallicity, radiation hardness, ionization, and heating from the old stellar population \citep{Lee2013,Remy-Ruyer2015,Shivaei2017,Lin2020,Mallory2022,Zhang2022,Zhang2022b,Zhang2022c}.

\subsubsection{Removing dependence of properties on SFR}
When studying the secondary dependence as a SFR indicator, 12 $\mu \rm m$ luminosity is typically normalized by (compared to) the TIR luminosity $\LTIR$ \citep[e.g.,][]{Cluver2017}, for the convenience of the same physical units. 
TIR flux comes from the dust-processed starlight that was originally in the UV and optical. There can be a non-negligible fraction of the TIR light absorbed by dust grains from the old stellar populations of galaxies \citep[e.g.,][]{Hirashita2003,DeLooze2014}. Thus the ratio of 12 $\mu m$ luminosity over TIR luminosity has a complex underlying dependence on the stellar population. 

In this work, we normalize the 12 $\mu \rm m$ luminosity by the $\mdust$ instead of by $\LTIR$, and meanwhile control for the SFR. The dust mass here when associated with SFR, can be viewed as proxy for the part of TIR that is directly related to SFR (instead of old stars). In the following, we use this methodology, and study the dependence of $\LWthreeMdust$ on galactic properties other than SFR. We mainly use the xCOLD GASS disk sample, and all scaling relations and correlation coefficients are derived based on it. We also over-plot the distribution of JINGLE disk galaxies in figures for reference.

\begin{figure}[htbp]
        \centering
	\includegraphics[width=1\linewidth]{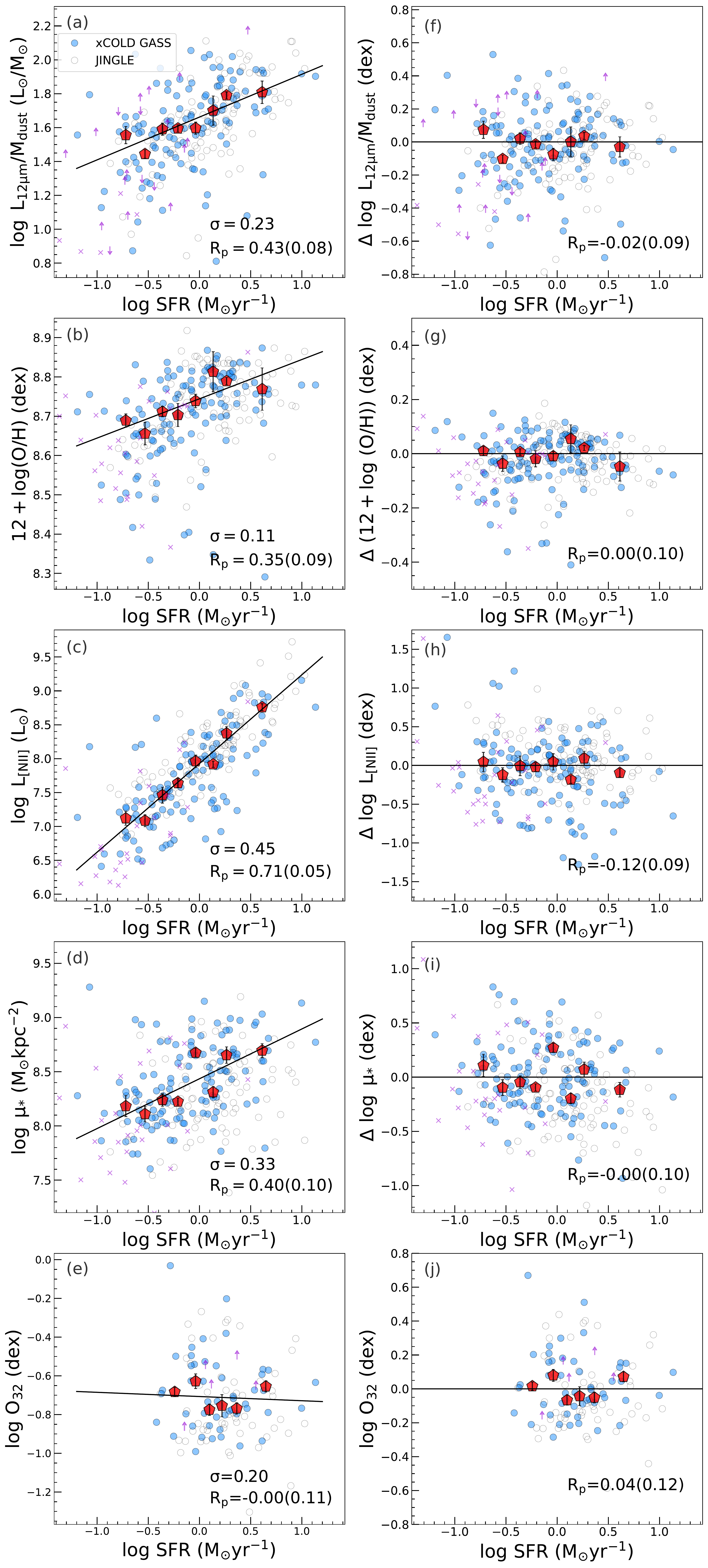}
	\caption{Reference relations for deriving the deviation of the galaxies. From panel (a) to (e), the large red pentagons indicate the medians of each parameter in SFR bins for xCOLD GASS disk sample with errors determined by bootstrapping, and the black lines show the best-fit least-squares regression lines. The open circles mark the JINGLE disk sample for reference. The scatter ($\sigma$), the Pearson's correlation coefficient ($R_{p}$, and its uncertainty), and Kendall’s $\tau$ coefficient ($R_{k}$, and its uncertainty) are listed in the lower right corner of each panel. From panel (f) to (j), we show the dependence of residuals from the best-fit relation on SFR.}
	\label{fig:SFR_relation_all.pdf}
\end{figure}

In order to control for the dependence on SFR, we calculate for each galactic property the residual from its scaling relations with SFR. First of all, we derive the scaling relations between $\LWthreeMdust$ and SFR shown in panel (a) of Figure \ref{fig:SFR_relation_all.pdf}. The xCOLD GASS and JINGLE galaxies are shown in blue and open circles, respectively. When deriving the scaling relation, we divide the sample into several SFR bins and then obtain the median SFR and median $\LWthreeMdust$ in each bin. The errorbars are derived through bootstrapping. Then we fit a linear line with these median values. The vertical distances of data points from the relation ($\Delta (\LWthreeMdust)$) are calculated, and are taken as residual from the scaling relations. 
We similarly derive residuals from scaling relation with SFR for other galactic properties of interest. These properties include gas-phase metallicity ($\rm 12 + \log(O/H)$), $\Nii$ line luminosity ($\LNII$), stellar mass surface density ($\mu_* = M_*(2\pi r^{2}_{50,z})^{-1}$) and the ionization parameter ($\oiii/\oii$, $O_{32}$, \citealt{Kewley2002,Morisset2016,Kewley2019}). 
 These relations and best-fit lines are displayed in panel (b) to (e) of Figure \ref{fig:SFR_relation_all.pdf}. It is worth noting that $\oii\ \lambda \lambda3727,29$ can only be observed in galaxies with redshift $z>0.02$ with SDSS spectrograph, therefore, we only consider a sub-sample containing 49 galaxies in xCOLD GASS disk sample to study the relation with $O_{32}$. 
 We plot the residuals from these scaling relations as a function of SFR in panel (f) to (j) of Figure \ref{fig:SFR_relation_all.pdf}. The low values of correlation coefficient in each panel show that the residuals from the relations do not have significant correlations with SFR, thus the secondary dependence on SFR has been successfully removed.

\subsubsection{Relations between $\Delta (\LWthreeMdust)$ and other residual properties}
\begin{figure*}[htbp]
	\includegraphics[width=\linewidth]{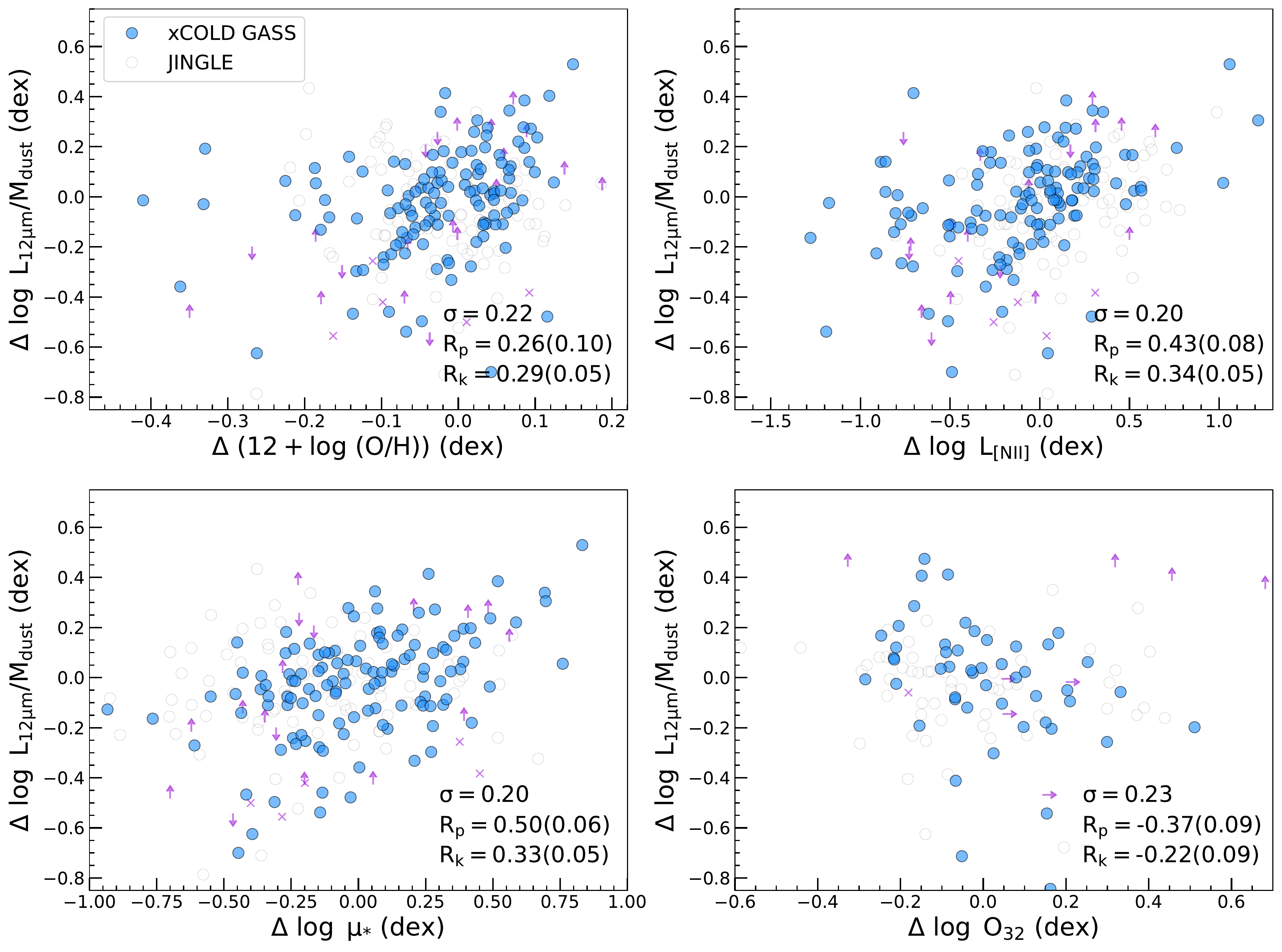}
	\caption{Correlations of $\Delta (\LWthreeMdust$) with $\Delta (\rm 12 + \log(O/H))$, $\Delta \LNII$, $\Delta \mu_{*}$ and $\Delta O_{32}$ for xCOLD GASS (in blue). The open circles mark the JINGLE disk sample. The $\Delta$ quantifies the residuals from the relation of each parameters with SFR to remove the dependence of SFR. The scatter ($\sigma$), the Pearson's correlation coefficient ($R_{p}$, and its uncertainty), and the Kendall’s $\tau$ coefficient ($R_{k}$, and its uncertainty) are listed in the lower right corner of each panel.}
	\label{fig:L12_Mdust_all.pdf}
\end{figure*}

We study the relations of $\Delta (\LWthreeMdust)$ versus $\Delta (\rm 12 + \log(O/H))$, $\Delta \LNII$, $\Delta \mu_*$, and $\Delta O_{32}$. The results are shown in Figure \ref{fig:L12_Mdust_all.pdf}. We summarize all the correlation coefficients in Table \ref{tab:all_relation}. We find that $\Delta (\LWthreeMdust)$ has a strong correlation with $\Delta \mu_*$, and moderate correlation with $\Delta \LNII$. It also has noticeable but weaker correlation with $\Delta (\rm 12 + \log(O/H))$ and $\Delta O_{32}$ than the correlation with $\LNII$.

\begin{deluxetable*}{ccccc}
\tablecaption{The Pearson's correlation coefficients and Kendall’s $\tau$ coefficients for the relations between $\Delta (\LWthreeMdust)$ and $\Delta X$ for xCOLD GASS disk sample.\label{tab:all_relation}}
\centering
\tabcolsep=12pt
\tablehead{
\colhead{y} & \colhead{x} & \colhead{Sample $^{a}$} & \colhead{Pearson's correlation coefficient}  & \colhead{Kendall’s $\tau$ coefficient}\\
}
\startdata
$\Delta (\LWthreeMdust)$ & $\Delta (\rm 12 + \log(O/H))$ & 125(20)  & $0.26\pm 0.10$ & $0.29\pm 0.05$\\
$\Delta (\LWthreeMdust)$ & $\Delta \LNII$ & 125(20) & $0.43\pm0.08$ & $0.34\pm0.05$\\
$\Delta (\LWthreeMdust)$ & $\Delta \mu_{*}$ & 125(20) & $0.50\pm0.06$ & $0.33\pm0.05$\\
$\Delta (\LWthreeMdust)$ & $\Delta \rm{O_{32}}$ & 49(8) & $-0.37\pm0.09$ & $-0.22\pm0.09$\\
\hline
\enddata
\begin{flushleft}
Notes. $^{a}$The value shows the number of galaxies detected in both 12 $\rm\mu m$ and gas. The value in parentheses represents the number of galaxies that are not detected by either 12 $\rm \mu m$ and gas.
\end{flushleft}

\end{deluxetable*}

PAH features are possibly substantially suppressed in low-metallicity and highly ionized environments \citep[e.g.,][]{Madden2006,Hunt2010,Xie2019,Li2020}. Metals promote the formation of PAH as catalysts \citep[e.g.,][]{Sandstrom2012}, and shield PAH from hard and strong UV radiation \citep[e.g.,][]{Hunt2011}. \citet{Shivaei2017} find 7.7 $\mu m$ PAH emission shows a significant anti-correlation with ionization sensitive parameter $O_{32}$, and this trend sometimes can be even stronger than the correlation with metallicity. 
After removing the effects of SFR (thus normalization of $H\alpha$), high values of $\Delta \LNII$ become an indicator of high metallicity and low ionization \citep[e.g.,][]{Denicolo2002,Pettini2004}. The dependence of $\Delta (\LWthreeMdust)$ on $\Delta \LNII$ may reflect the aforementioned scenario of PAH being formed and destroyed. The better correlation of $\Delta (\LWthreeMdust)$ with $\Delta \LNII$ than with $\Delta (\rm 12 + \log(O/H))$ or $\Delta O_{32}$ implies that $\Delta \LNII$ better captures the scenario than metallicity or ionization parameter alone.

The correlation between $\Delta (\LWthreeMdust)$ and $\Delta \mu_{*}$ 
is consistent with the recent finding that on kpc scales and at a fixed SFR, higher stellar mass is associated with stronger PAH emission \citep{Zhang2022}. The interpretation is that light from evolved stars is likely to contribute to heating the PAH. We further test and find that the correlation with $\Delta (\LWthreeMdust)$ is stronger for $\Delta \mu_{*}$ than for $\Delta M_{*}$, possibly because stellar mass on kpc scales is physically closer to $\mu_{*}$ than to the integral $M_{*}$. The result of this comparison also eliminates the possibility that the residual dependence on $M_{*}$ when predicting the $\mdust$ with $\mgasin$ (Section \ref{subsec:residual}) could be the cause of the relationship between $\Delta (\LWthreeMdust)$ and $\mu_*$.

We briefly comment that, if the JINGLE disk sample is included in the analysis, then all the relations of $\Delta (\LWthreeMdust)$ in Figure \ref{fig:L12_Mdust_all.pdf} become weaker. This change in results imply the caution needed when analyzing a biased sample.

In summary, the dependence of integral 12 $\mu m$ luminosity on integral galactic properties investigated here based on a relatively representative sample of disk galaxies shows a high level of consistency with that found on kpc scales and from more detailed mapping-mode spectroscopy and SED modeling. The integral 12 $\mu m$ luminosity shows dependence on metallicity, ionization, and the stellar mass surface density, and thus can not be used as a simple SFR indicator without correction for the secondary dependence. 

\subsection{Summary of the application to xCOLD GASS and future Perspective}\label{sec:discussion}

We have shown that the dust mass scaling relations derived for the xCOLD GASS disk sample are found to be in agreement with previous studies. We have also confirmed that, when controlling for the dust mass, previously known secondary dependence of PAH emission on metallicity, ionization, and mass of the evolved stellar population on kpc-scales (based on spectroscopy) remains on global scales.
These consistencies suggest that the application of the inner gas--dust mass relation to derive the dust mass for disk galaxies from the xCOLD GASS sample is relatively successful. 

Thus, the dust--inner gas mass relation provides a promising tool to derive $\mgasin$ from $\mdust$ or the other way round for disk galaxies. 
Here, for a future perspective, we highlight the first type of conversion because of the long-last difficulty to directly observe the cold gas. 
Although SKA path finders \citep{Koribalski2020} and FAST \citep{Li2016} have launched big surveys to map $\hi$ for practically the whole sky, the intermediate- to high-redshift universe, as well as the gas-poor regime of galaxy population, will continue to pose challenges for $\hi$ observations. 
Wide-field $\htwo$ (or CO) surveys are still largely missing , so $\htwo$ is often indicated by dust observations \citep[e.g.,][]{Eales2012,Genzel2015,Groves2015,Janowiecki2018} or optical spectroscopy data \citep[e.g.,][]{Yesuf2019}. We have shown that at least for disk galaxies the dust mass (typically detected within the optical radius) is more closely related to the inner gas mass than the integral gas mass or the molecular gas mass alone. Theoretically the inner gas mass should provide useful clues to galaxy evolution as it is the immediately reservoir of material for star formation \citep{Wang2020,Chen2022}, but its role has been poorly studied so far due to the previous sparsity of measurements. 
The strong relation of dust mass with inner gas mass provides the opportunity and motivation for us to better characterize and understand the role of inner gas mass in disk galaxies. 
Basing on the observations of dust (mass or FIR luminosity) with IR instruments (e.g., Origins Space Telescope, \citealt{Meixner2019}) or submillimeter instruments for high-redshift galaxies (e.g., ALMA, \citealt{Gonzalez-Lopez2017,Faisst2020,Fudamoto2022}; JCMT, \citealt{Zavala2015}; AtLAST, \citealt{Ramasawmy2022}), we can obtain more accurate $\mgasin$ than before for disk galaxies in a large volume of the universe, as well as for the gas-poor disk galaxies. 

\section{Summary}\label{sec:summary}
Using a sample of disk-like galaxies from HRS, we have studied the relationship between gas-to-dust ratio and metallicity in the inner region (within the optical $\rnine$) of galaxies. We find the following:

\begin{enumerate}
\item Compared to $\xigas$, the $\xigasin$ shows much less significant correlation with the metallicity. It indicates that, at least part of the relation between $\xigas$ and metallicity is due to the inclusion of metal-poor and outlying $\hi$ gas. Compared to $\xihtwo$, the $\xigasin$ shows much smaller scatter at a given metallicity. It confirms that the total gas mass is a better tracer of dust mass than the molecular gas mass alone, both for the gas in the inner disk and on global scale.

\item The inner gas mass shows a stronger and tighter correlation with the dust mass than that of integral gas mass. The correlation can be even stronger, if we replace the dust mass with $\Lfive$. These strong connections enable us to better predict the inner gas mass based on IR$\slash$dust measurements, or the other way round.

\item We calibrate scaling relations between the inner gas mass, dust mass and $\Lfive$. Depending on the relation in consideration, there is only weak dependence of the residual on the stellar mass, metallicity, or SFR. 

\end{enumerate}

We demonstrate a simple application of the scaling relations derived above, by predicting dust mass based on the inner gas mass for disk-like galaxies from xCOLD GASS. Our main results are the following: 

\begin{enumerate}

\item The predicted specific dust masses show scaling relations with $\mstar$ and with sSFR. These relations are consistent with the fiducial relations in the literature, which supports that the $\mgasin$--$\mdust$ relation could provide a reasonable prediction of dust mass.

\item After controlling for the dust mass and SFR, the WISE W3-band luminosity $\LWthree$ strongly  depends on the stellar mass surface density which is an indicator of contribution to PAH heating from evolved stars. It also depends on the $\Nii$ luminosity which is a combined indicator of both metallicity and ionization. 
These secondary correlations imply the danger of directly using $\LWthree$ as SFR indicator on galactic global scales. They are consistent with previous conclusion based on more detailed analysis of mapping-mode MIR spectroscopy of nearby individual galaxies \citep[e.g.,][]{Lee2013,Gregg2022}. 

\end{enumerate}

\section*{Acknowledgements}
We thank all the people for useful discussions. This work is supported by the Strategic Priority Research Program of Chinese Academy of Sciences (Grant No. XDB 41000000), the National Science Foundation of China (NSFC, Grant No. 12233008, 11973038), the China Manned Space Project (No. CMS-CSST-2021-A07) and the Cyrus Chun Ying Tang Foundations. JW thank support of research grants from National Science Foundation of China (NO. 12073002, 12233001), Ministry of Science and Technology of the People's Republic of China (NO. 2022YFA1602902), and the China Manned Space Project (NO. CMS-CSST-2021-B02).

This work has made use of Python (\href{http://www.python.org}{http://www.python.org}) and the Python packages: astropy \citep{Astropy2013}, NumPy (\href{http://www.numpy.org/}{http://www.numpy.org/}), matplotlib (\href{http://www.numpy.org/}{https://matplotlib.org/}) \citep{Hunter2007} and photutils \citep{Bradley2022}. Furthermore NADA package in R programming language has been used (\href{https://CRAN.R-project.org/package=NADA}{https://CRAN.R-project.org/package=NADA}).

This research has made use of the VizieR catalogue access tool, CDS, Strasbourg, France \citep{Ochsenbein2000}. This research has also made use of the NASA/IPAC Infrared Science Archive, which is operated by the Jet Propulsion Laboratory, California Institute of Technology, under contract with the National Aeronautics and Space Administration.

\bibliography{inner_GTD}

\begin{thebibliography}{}
\expandafter\ifx\csname natexlab\endcsname\relax\def\natexlab#1{#1}\fi
\providecommand{\url}[1]{\href{#1}{#1}}
\providecommand{\dodoi}[1]{doi:~\href{http://doi.org/#1}{\nolinkurl{#1}}}
\providecommand{\doeprint}[1]{\href{http://ascl.net/#1}{\nolinkurl{http://ascl.net/#1}}}
\providecommand{\doarXiv}[1]{\href{https://arxiv.org/abs/#1}{\nolinkurl{https://arxiv.org/abs/#1}}}

\bibitem[{{Abazajian} {et~al.}(2009){Abazajian}, {Adelman-McCarthy},
  {Ag{\"u}eros}, {Allam}, {Allende Prieto}, {An}, {Anderson}, {Anderson},
  {Annis}, {Bahcall}, {Bailer-Jones}, {Barentine}, {Bassett}, {Becker},
  {Beers}, {Bell}, {Belokurov}, {Berlind}, {Berman}, {Bernardi}, {Bickerton},
  {Bizyaev}, {Blakeslee}, {Blanton}, {Bochanski}, {Boroski}, {Brewington},
  {Brinchmann}, {Brinkmann}, {Brunner}, {Budav{\'a}ri}, {Carey}, {Carliles},
  {Carr}, {Castander}, {Cinabro}, {Connolly}, {Csabai}, {Cunha}, {Czarapata},
  {Davenport}, {de Haas}, {Dilday}, {Doi}, {Eisenstein}, {Evans}, {Evans},
  {Fan}, {Friedman}, {Frieman}, {Fukugita}, {G{\"a}nsicke}, {Gates},
  {Gillespie}, {Gilmore}, {Gonzalez}, {Gonzalez}, {Grebel}, {Gunn},
  {Gy{\"o}ry}, {Hall}, {Harding}, {Harris}, {Harvanek}, {Hawley}, {Hayes},
  {Heckman}, {Hendry}, {Hennessy}, {Hindsley}, {Hoblitt}, {Hogan}, {Hogg},
  {Holtzman}, {Hyde}, {Ichikawa}, {Ichikawa}, {Im}, {Ivezi{\'c}}, {Jester},
  {Jiang}, {Johnson}, {Jorgensen}, {Juri{\'c}}, {Kent}, {Kessler}, {Kleinman},
  {Knapp}, {Konishi}, {Kron}, {Krzesinski}, {Kuropatkin}, {Lampeitl},
  {Lebedeva}, {Lee}, {Lee}, {French Leger}, {L{\'e}pine}, {Li}, {Lima}, {Lin},
  {Long}, {Loomis}, {Loveday}, {Lupton}, {Magnier}, {Malanushenko},
  {Malanushenko}, {Mandelbaum}, {Margon}, {Marriner}, {Mart{\'\i}nez-Delgado},
  {Matsubara}, {McGehee}, {McKay}, {Meiksin}, {Morrison}, {Mullally}, {Munn},
  {Murphy}, {Nash}, {Nebot}, {Neilsen}, {Newberg}, {Newman}, {Nichol},
  {Nicinski}, {Nieto-Santisteban}, {Nitta}, {Okamura}, {Oravetz}, {Ostriker},
  {Owen}, {Padmanabhan}, {Pan}, {Park}, {Pauls}, {Peoples}, {Percival}, {Pier},
  {Pope}, {Pourbaix}, {Price}, {Purger}, {Quinn}, {Raddick}, {Re Fiorentin},
  {Richards}, {Richmond}, {Riess}, {Rix}, {Rockosi}, {Sako}, {Schlegel},
  {Schneider}, {Scholz}, {Schreiber}, {Schwope}, {Seljak}, {Sesar}, {Sheldon},
  {Shimasaku}, {Sibley}, {Simmons}, {Sivarani}, {Allyn Smith}, {Smith},
  {Smol{\v{c}}i{\'c}}, {Snedden}, {Stebbins}, {Steinmetz}, {Stoughton},
  {Strauss}, {SubbaRao}, {Suto}, {Szalay}, {Szapudi}, {Szkody}, {Tanaka},
  {Tegmark}, {Teodoro}, {Thakar}, {Tremonti}, {Tucker}, {Uomoto}, {Vanden
  Berk}, {Vandenberg}, {Vidrih}, {Vogeley}, {Voges}, {Vogt}, {Wadadekar},
  {Watters}, {Weinberg}, {West}, {White}, {Wilhite}, {Wonders}, {Yanny},
  {Yocum}, {York}, {Zehavi}, {Zibetti}, \& {Zucker}}]{Abazajian2009}
{Abazajian}, K.~N., {Adelman-McCarthy}, J.~K., {Ag{\"u}eros}, M.~A., {et~al.}
  2009, \apjs, 182, 543, \dodoi{10.1088/0067-0049/182/2/543}

\bibitem[{{Abdurro'uf} {et~al.}(2022{\natexlab{a}}){Abdurro'uf}, {Lin},
  {Hirashita}, {Morishita}, {Tacchella}, {Akiyama}, {Takeuchi}, \&
  {Wu}}]{Abdurro'uf2022}
{Abdurro'uf}, {Lin}, Y.-T., {Hirashita}, H., {et~al.} 2022{\natexlab{a}}, \apj,
  926, 81, \dodoi{10.3847/1538-4357/ac439a}

\bibitem[{{Abdurro'uf} {et~al.}(2022{\natexlab{b}}){Abdurro'uf}, {Lin},
  {Hirashita}, {Morishita}, {Tacchella}, {Wu}, {Akiyama}, \&
  {Takeuchi}}]{Abdurro'uf2022b}
---. 2022{\natexlab{b}}, \apj, 935, 98, \dodoi{10.3847/1538-4357/ac7da4}

\bibitem[{Accurso {et~al.}(2017)Accurso, Saintonge, Catinella, Cortese,
  Dav{\'{e}}, Dunsheath, Genzel, Gracia-Carpio, Heckman, Jimmy, Kramer, Li,
  Lutz, Schiminovich, Schuster, Sternberg, Sturm, Tacconi, Tran, \&
  Wang}]{Accurso2017}
Accurso, G., Saintonge, A., Catinella, B., {et~al.} 2017, Monthly Notices of
  the Royal Astronomical Society, \dodoi{10.1093/mnras/stx1556}

\bibitem[{{Alton} {et~al.}(1998){Alton}, {Trewhella}, {Davies}, {Evans},
  {Bianchi}, {Gear}, {Thronson}, {Valentijn}, \& {Witt}}]{Alton1998}
{Alton}, P.~B., {Trewhella}, M., {Davies}, J.~I., {et~al.} 1998, \aap, 335, 807

\bibitem[{{Aniano} {et~al.}(2020){Aniano}, {Draine}, {Hunt}, {Sandstrom},
  {Calzetti}, {Kennicutt}, {Dale}, {Galametz}, {Gordon}, {Leroy}, {Smith},
  {Roussel}, {Sauvage}, {Walter}, {Armus}, {Bolatto}, {Boquien}, {Crocker}, {De
  Looze}, {Donovan Meyer}, {Helou}, {Hinz}, {Johnson}, {Koda}, {Miller},
  {Montiel}, {Murphy}, {Rela{\~n}o}, {Rix}, {Schinnerer}, {Skibba}, {Wolfire},
  \& {Engelbracht}}]{Aniano2020}
{Aniano}, G., {Draine}, B.~T., {Hunt}, L.~K., {et~al.} 2020, \apj, 889, 150,
  \dodoi{10.3847/1538-4357/ab5fdb}

\bibitem[{{Aravena} {et~al.}(2014){Aravena}, {Hodge}, {Wagg}, {Carilli},
  {Daddi}, {Dannerbauer}, {Lentati}, {Riechers}, {Sargent}, \&
  {Walter}}]{Aravena2014}
{Aravena}, M., {Hodge}, J.~A., {Wagg}, J., {et~al.} 2014, \mnras, 442, 558,
  \dodoi{10.1093/mnras/stu838}

\bibitem[{{Astropy Collaboration} {et~al.}(2013){Astropy Collaboration},
  {Robitaille}, {Tollerud}, {Greenfield}, {Droettboom}, {Bray}, {Aldcroft},
  {Davis}, {Ginsburg}, {Price-Whelan}, {Kerzendorf}, {Conley}, {Crighton},
  {Barbary}, {Muna}, {Ferguson}, {Grollier}, {Parikh}, {Nair}, {Unther},
  {Deil}, {Woillez}, {Conseil}, {Kramer}, {Turner}, {Singer}, {Fox}, {Weaver},
  {Zabalza}, {Edwards}, {Azalee Bostroem}, {Burke}, {Casey}, {Crawford},
  {Dencheva}, {Ely}, {Jenness}, {Labrie}, {Lim}, {Pierfederici}, {Pontzen},
  {Ptak}, {Refsdal}, {Servillat}, \& {Streicher}}]{Astropy2013}
{Astropy Collaboration}, {Robitaille}, T.~P., {Tollerud}, E.~J., {et~al.} 2013,
  \aap, 558, A33, \dodoi{10.1051/0004-6361/201322068}

\bibitem[{{Baldwin} {et~al.}(1981){Baldwin}, {Phillips}, \&
  {Terlevich}}]{Baldwin1981}
{Baldwin}, J.~A., {Phillips}, M.~M., \& {Terlevich}, R. 1981, \pasp, 93, 5,
  \dodoi{10.1086/130766}

\bibitem[{Belfiore {et~al.}(2017)Belfiore, Maiolino, Tremonti, S{\'{a}}nchez,
  Bundy, Bershady, Westfall, Lin, Drory, Boquien, Thomas, \&
  Brinkmann}]{Belfiore2017}
Belfiore, F., Maiolino, R., Tremonti, C., {et~al.} 2017, Monthly Notices of the
  Royal Astronomical Society, 469, 151, \dodoi{10.1093/mnras/stx789}

\bibitem[{Berta {et~al.}(2016)Berta, Lutz, Genzel, F\"{o}rster-Schreiber, \&
  Tacconi}]{Berta2016}
Berta, S., Lutz, D., Genzel, R., F\"{o}rster-Schreiber, N.~M., \& Tacconi,
  L.~J. 2016, 587, A73, \dodoi{10.1051/0004-6361/201527746}

\bibitem[{{Bertemes} {et~al.}(2018){Bertemes}, {Wuyts}, {Lutz}, {F{\"o}rster
  Schreiber}, {Genzel}, {Minchin}, {Mundell}, {Rosario}, {Saintonge}, \&
  {Tacconi}}]{Bertemes2018}
{Bertemes}, C., {Wuyts}, S., {Lutz}, D., {et~al.} 2018, \mnras, 478, 1442,
  \dodoi{10.1093/mnras/sty963}

\bibitem[{{Bianchi}(2007)}]{Bianchi2007}
{Bianchi}, S. 2007, \aap, 471, 765, \dodoi{10.1051/0004-6361:20077649}

\bibitem[{{Bigiel} {et~al.}(2008){Bigiel}, {Leroy}, {Walter}, {Brinks}, {de
  Blok}, {Madore}, \& {Thornley}}]{Bigiel2008}
{Bigiel}, F., {Leroy}, A., {Walter}, F., {et~al.} 2008, \aj, 136, 2846,
  \dodoi{10.1088/0004-6256/136/6/2846}

\bibitem[{{Blanton} {et~al.}(2011){Blanton}, {Kazin}, {Muna}, {Weaver}, \&
  {Price-Whelan}}]{Blanton2011}
{Blanton}, M.~R., {Kazin}, E., {Muna}, D., {Weaver}, B.~A., \& {Price-Whelan},
  A. 2011, \aj, 142, 31, \dodoi{10.1088/0004-6256/142/1/31}

\bibitem[{{Boquien} \& {Salim}(2021)}]{Boquien2021}
{Boquien}, M., \& {Salim}, S. 2021, \aap, 653, A149,
  \dodoi{10.1051/0004-6361/202140992}

\bibitem[{{Boselli} {et~al.}(2014){Boselli}, {Cortese}, \&
  {Boquien}}]{Boselli2014}
{Boselli}, A., {Cortese}, L., \& {Boquien}, M. 2014, \aap, 564, A65,
  \dodoi{10.1051/0004-6361/201322311}

\bibitem[{{Boselli} {et~al.}(2015){Boselli}, {Fossati}, {Gavazzi}, {Ciesla},
  {Buat}, {Boissier}, \& {Hughes}}]{Boselli2015}
{Boselli}, A., {Fossati}, M., {Gavazzi}, G., {et~al.} 2015, \aap, 579, A102,
  \dodoi{10.1051/0004-6361/201525712}

\bibitem[{{Boselli} {et~al.}(2002){Boselli}, {Lequeux}, \&
  {Gavazzi}}]{Boselli2002}
{Boselli}, A., {Lequeux}, J., \& {Gavazzi}, G. 2002, \aap, 384, 33,
  \dodoi{10.1051/0004-6361:20011747}

\bibitem[{Boselli {et~al.}(2010)Boselli, Eales, Cortese, Bendo, Chanial, Buat,
  Davies, Auld, Rigby, Baes, Barlow, Bock, Bradford, Castro-Rodriguez, Charlot,
  Clements, Cormier, Dwek, Elbaz, Galametz, Galliano, Gear, Glenn, Gomez,
  Griffin, Hony, Isaak, Levenson, Lu, Madden, O'Halloran, Okamura, Oliver,
  Page, Panuzzo, Papageorgiou, Parkin, Perez-Fournon, Pohlen, Rangwala,
  Roussel, Rykala, Sacchi, Sauvage, Schulz, Schirm, Smith, Spinoglio, Stevens,
  Symeonidis, Vaccari, Vigroux, Wilson, Wozniak, Wright, \&
  Zeilinger}]{Boselli2010}
Boselli, A., Eales, S., Cortese, L., {et~al.} 2010, Publications of the
  Astronomical Society of the Pacific, 122, 261, \dodoi{10.1086/651535}

\bibitem[{Bradley {et~al.}(2019)Bradley, Sipőcz, Robitaille, Tollerud, {Zé
  Vinícius}, Deil, Barbary, Wilson, Busko, G\"{u}nther, Cara, Conseil,
  Droettboom, {Azalee Bostroem}, {E. M. Bray}, Bratholm, {P. L. Lim}, Craig,
  Barentsen, Pascual, Donath, Greco, Perren, Kerzendorf, Val-Borro, Dencheva,
  Ferreira, Souchereau, D'Eugenio, \& Weaver}]{Bradley2019}
Bradley, L., Sipőcz, B., Robitaille, T., {et~al.} 2019, astropy/photutils:
  v0.7.2,  Zenodo, \dodoi{10.5281/ZENODO.3568287}

\bibitem[{Bradley {et~al.}(2022)Bradley, Sipőcz, Robitaille, Tollerud,
  Vinícius, Deil, Barbary, Wilson, Busko, Donath, Günther, Cara, Lim,
  Meßlinger, Conseil, Bostroem, Droettboom, Bray, Bratholm, Barentsen, Craig,
  Rathi, Pascual, Perren, Georgiev, de~Val-Borro, Kerzendorf, Bach, Quint, \&
  Souchereau}]{Bradley2022}
---. 2022, astropy/photutils: 1.5.0, 1.5.0,  Zenodo,
  \dodoi{10.5281/zenodo.6825092}

\bibitem[{{Broeils} \& {Rhee}(1997)}]{Broeils1997}
{Broeils}, A.~H., \& {Rhee}, M.~H. 1997, \aap, 324, 877

\bibitem[{{Brown} {et~al.}(2014){Brown}, {Jarrett}, \& {Cluver}}]{Brown2014}
{Brown}, M.~J.~I., {Jarrett}, T.~H., \& {Cluver}, M.~E. 2014, \pasa, 31, e049,
  \dodoi{10.1017/pasa.2014.44}

\bibitem[{{Calura} {et~al.}(2017){Calura}, {Pozzi}, {Cresci}, {Santini},
  {Gruppioni}, {Pozzetti}, {Gilli}, {Matteucci}, \& {Maiolino}}]{Calura2017}
{Calura}, F., {Pozzi}, F., {Cresci}, G., {et~al.} 2017, \mnras, 465, 54,
  \dodoi{10.1093/mnras/stw2749}

\bibitem[{{Calzetti} {et~al.}(2000){Calzetti}, {Armus}, {Bohlin}, {Kinney},
  {Koornneef}, \& {Storchi-Bergmann}}]{Calzetti2000}
{Calzetti}, D., {Armus}, L., {Bohlin}, R.~C., {et~al.} 2000, \apj, 533, 682,
  \dodoi{10.1086/308692}

\bibitem[{{Casasola} {et~al.}(2017){Casasola}, {Cassar{\`a}}, {Bianchi},
  {Verstocken}, {Xilouris}, {Magrini}, {Smith}, {De Looze}, {Galametz},
  {Madden}, {Baes}, {Clark}, {Davies}, {De Vis}, {Evans}, {Fritz}, {Galliano},
  {Jones}, {Mosenkov}, {Viaene}, \& {Ysard}}]{Casasola2017}
{Casasola}, V., {Cassar{\`a}}, L.~P., {Bianchi}, S., {et~al.} 2017, \aap, 605,
  A18, \dodoi{10.1051/0004-6361/201731020}

\bibitem[{{Casasola} {et~al.}(2020){Casasola}, {Bianchi}, {De Vis}, {Magrini},
  {Corbelli}, {Clark}, {Fritz}, {Nersesian}, {Viaene}, {Baes}, {Cassar{\`a}},
  {Davies}, {De Looze}, {Dobbels}, {Galametz}, {Galliano}, {Jones}, {Madden},
  {Mosenkov}, {Tr{\v{c}}ka}, \& {Xilouris}}]{Casasola2020}
{Casasola}, V., {Bianchi}, S., {De Vis}, P., {et~al.} 2020, \aap, 633, A100,
  \dodoi{10.1051/0004-6361/201936665}

\bibitem[{{Casasola} {et~al.}(2022){Casasola}, {Bianchi}, {Magrini},
  {Mosenkov}, {Salvestrini}, {Baes}, {Calura}, {Cassar{\`a}}, {Clark},
  {Corbelli}, {Fritz}, {Galliano}, {Liuzzo}, {Madden}, {Nersesian}, {Pozzi},
  {Roychowdhury}, {Baronchelli}, {Bonato}, {Gruppioni}, \&
  {Pantoni}}]{Casasola2022}
{Casasola}, V., {Bianchi}, S., {Magrini}, L., {et~al.} 2022, \aap, 668, A130,
  \dodoi{10.1051/0004-6361/202245043}

\bibitem[{{Catinella} {et~al.}(2018){Catinella}, {Saintonge}, {Janowiecki},
  {Cortese}, {Dav{\'e}}, {Lemonias}, {Cooper}, {Schiminovich}, {Hummels},
  {Fabello}, {Ger{\'e}b}, {Kilborn}, \& {Wang}}]{Catinella2018}
{Catinella}, B., {Saintonge}, A., {Janowiecki}, S., {et~al.} 2018, \mnras, 476,
  875, \dodoi{10.1093/mnras/sty089}

\bibitem[{{Chabrier}(2003)}]{Chabrier2003}
{Chabrier}, G. 2003, \pasp, 115, 763, \dodoi{10.1086/376392}

\bibitem[{{Chen} {et~al.}(2022){Chen}, {Wang}, \& {Kong}}]{Chen2022}
{Chen}, X., {Wang}, J., \& {Kong}, X. 2022, \apj, 933, 39,
  \dodoi{10.3847/1538-4357/ac70d0}

\bibitem[{{Ciesla} {et~al.}(2012){Ciesla}, {Boselli}, {Smith}, {Bendo},
  {Cortese}, {Eales}, {Bianchi}, {Boquien}, {Buat}, {Davies}, {Pohlen},
  {Zibetti}, {Baes}, {Cooray}, {De Looze}, {di Serego Alighieri}, {Galametz},
  {Gomez}, {Lebouteiller}, {Madden}, {Pappalardo}, {Remy}, {Spinoglio},
  {Vaccari}, {Auld}, \& {Clements}}]{Ciesla2012}
{Ciesla}, L., {Boselli}, A., {Smith}, M.~W.~L., {et~al.} 2012, \aap, 543, A161,
  \dodoi{10.1051/0004-6361/201219216}

\bibitem[{{Ciesla} {et~al.}(2014){Ciesla}, {Boquien}, {Boselli}, {Buat},
  {Cortese}, {Bendo}, {Heinis}, {Galametz}, {Eales}, {Smith}, {Baes},
  {Bianchi}, {De Looze}, {di Serego Alighieri}, {Galliano}, {Hughes}, {Madden},
  {Pierini}, {R{\'e}my-Ruyer}, {Spinoglio}, {Vaccari}, {Viaene}, \&
  {Vlahakis}}]{Ciesla2014}
{Ciesla}, L., {Boquien}, M., {Boselli}, A., {et~al.} 2014, \aap, 565, A128,
  \dodoi{10.1051/0004-6361/201323248}

\bibitem[{{Clark} {et~al.}(2015){Clark}, {Dunne}, {Gomez}, {Maddox}, {De Vis},
  {Smith}, {Eales}, {Baes}, {Bendo}, {Bourne}, {Driver}, {Dye}, {Furlanetto},
  {Grootes}, {Ivison}, {Schofield}, {Robotham}, {Rowlands}, {Valiante},
  {Vlahakis}, {van der Werf}, {Wright}, \& {de Zotti}}]{Clark2015}
{Clark}, C.~J.~R., {Dunne}, L., {Gomez}, H.~L., {et~al.} 2015, \mnras, 452,
  397, \dodoi{10.1093/mnras/stv1276}

\bibitem[{{Clemens} {et~al.}(2013){Clemens}, {Negrello}, {De Zotti},
  {Gonzalez-Nuevo}, {Bonavera}, {Cosco}, {Guarese}, {Boaretto}, {Salucci},
  {Baccigalupi}, {Clements}, {Danese}, {Lapi}, {Mandolesi}, {Partridge},
  {Perrotta}, {Serjeant}, {Scott}, \& {Toffolatti}}]{Clemens2013}
{Clemens}, M.~S., {Negrello}, M., {De Zotti}, G., {et~al.} 2013, \mnras, 433,
  695, \dodoi{10.1093/mnras/stt760}

\bibitem[{Cluver {et~al.}(2017)Cluver, Jarrett, Dale, Smith, August, \&
  Brown}]{Cluver2017}
Cluver, M.~E., Jarrett, T.~H., Dale, D.~A., {et~al.} 2017, The Astrophysical
  Journal, 850, 68, \dodoi{10.3847/1538-4357/aa92c7}

\bibitem[{{Corbelli} {et~al.}(2012){Corbelli}, {Bianchi}, {Cortese},
  {Giovanardi}, {Magrini}, {Pappalardo}, {Boselli}, {Bendo}, {Davies},
  {Grossi}, {Madden}, {Smith}, {Vlahakis}, {Auld}, {Baes}, {De Looze}, {Fritz},
  {Pohlen}, \& {Verstappen}}]{Corbelli2012}
{Corbelli}, E., {Bianchi}, S., {Cortese}, L., {et~al.} 2012, \aap, 542, A32,
  \dodoi{10.1051/0004-6361/201117329}

\bibitem[{Cortese {et~al.}(2012)Cortese, Ciesla, Boselli, Bianchi, Gomez,
  Smith, Bendo, Eales, Pohlen, Baes, Corbelli, Davies, Hughes, Hunt, Madden,
  Pierini, di~Serego~Alighieri, Zibetti, Boquien, Clements, Cooray, Galametz,
  Magrini, Pappalardo, Spinoglio, \& Vlahakis}]{Cortese2012}
Cortese, L., Ciesla, L., Boselli, A., {et~al.} 2012, Astronomy {\&}
  Astrophysics, 540, A52, \dodoi{10.1051/0004-6361/201118499}

\bibitem[{{Cortese} {et~al.}(2012){Cortese}, {Ciesla}, {Boselli}, {Bianchi},
  {Gomez}, {Smith}, {Bendo}, {Eales}, {Pohlen}, {Baes}, {Corbelli}, {Davies},
  {Hughes}, {Hunt}, {Madden}, {Pierini}, {di Serego Alighieri}, {Zibetti},
  {Boquien}, {Clements}, {Cooray}, {Galametz}, {Magrini}, {Pappalardo},
  {Spinoglio}, \& {Vlahakis}}]{Cortese2012b}
{Cortese}, L., {Ciesla}, L., {Boselli}, A., {et~al.} 2012, \aap, 540, A52,
  \dodoi{10.1051/0004-6361/201118499}

\bibitem[{{da Cunha} {et~al.}(2010){da Cunha}, {Eminian}, {Charlot}, \&
  {Blaizot}}]{daCunha2010}
{da Cunha}, E., {Eminian}, C., {Charlot}, S., \& {Blaizot}, J. 2010, \mnras,
  403, 1894, \dodoi{10.1111/j.1365-2966.2010.16344.x}

\bibitem[{{Daddi} {et~al.}(2010){Daddi}, {Bournaud}, {Walter}, {Dannerbauer},
  {Carilli}, {Dickinson}, {Elbaz}, {Morrison}, {Riechers}, {Onodera}, {Salmi},
  {Krips}, \& {Stern}}]{Daddi2010}
{Daddi}, E., {Bournaud}, F., {Walter}, F., {et~al.} 2010, \apj, 713, 686,
  \dodoi{10.1088/0004-637X/713/1/686}

\bibitem[{{De Looze} {et~al.}(2014){De Looze}, {Fritz}, {Baes}, {Bendo},
  {Cortese}, {Boquien}, {Boselli}, {Camps}, {Cooray}, {Cormier}, {Davies}, {De
  Geyter}, {Hughes}, {Jones}, {Karczewski}, {Lebouteiller}, {Lu}, {Madden},
  {R{\'e}my-Ruyer}, {Spinoglio}, {Smith}, {Viaene}, \& {Wilson}}]{DeLooze2014}
{De Looze}, I., {Fritz}, J., {Baes}, M., {et~al.} 2014, \aap, 571, A69,
  \dodoi{10.1051/0004-6361/201424747}

\bibitem[{{De Looze} {et~al.}(2020){De Looze}, {Lamperti}, {Saintonge},
  {Rela{\~n}o}, {Smith}, {Clark}, {Wilson}, {Decleir}, {Jones}, {Kennicutt},
  {Accurso}, {Brinks}, {Bureau}, {Cigan}, {Clements}, {De Vis}, {Fanciullo},
  {Gao}, {Gear}, {Ho}, {Hwang}, {Micha{\l}owski}, {Lee}, {Li}, {Lin}, {Liu},
  {Lomaeva}, {Pan}, {Sargent}, {Williams}, {Xiao}, \& {Zhu}}]{Looze2020}
{De Looze}, I., {Lamperti}, I., {Saintonge}, A., {et~al.} 2020, \mnras, 496,
  3668, \dodoi{10.1093/mnras/staa1496}

\bibitem[{{De Vis} {et~al.}(2017){De Vis}, {Dunne}, {Maddox}, {Gomez}, {Clark},
  {Bauer}, {Viaene}, {Schofield}, {Baes}, {Baker}, {Bourne}, {Driver}, {Dye},
  {Eales}, {Furlanetto}, {Ivison}, {Robotham}, {Rowlands}, {Smith}, {Smith},
  {Valiante}, \& {Wright}}]{DeVis2017}
{De Vis}, P., {Dunne}, L., {Maddox}, S., {et~al.} 2017, \mnras, 464, 4680,
  \dodoi{10.1093/mnras/stw2501}

\bibitem[{{De Vis} {et~al.}(2019){De Vis}, {Jones}, {Viaene}, {Casasola},
  {Clark}, {Baes}, {Bianchi}, {Cassara}, {Davies}, {De Looze}, {Galametz},
  {Galliano}, {Lianou}, {Madden}, {Manilla-Robles}, {Mosenkov}, {Nersesian},
  {Roychowdhury}, {Xilouris}, \& {Ysard}}]{DeVis2019}
{De Vis}, P., {Jones}, A., {Viaene}, S., {et~al.} 2019, \aap, 623, A5,
  \dodoi{10.1051/0004-6361/201834444}

\bibitem[{{Denicol{\'o}} {et~al.}(2002){Denicol{\'o}}, {Terlevich}, \&
  {Terlevich}}]{Denicolo2002}
{Denicol{\'o}}, G., {Terlevich}, R., \& {Terlevich}, E. 2002, \mnras, 330, 69,
  \dodoi{10.1046/j.1365-8711.2002.05041.x}

\bibitem[{{Draine} \& {Li}(2007)}]{Draine2007}
{Draine}, B.~T., \& {Li}, A. 2007, \apj, 657, 810, \dodoi{10.1086/511055}

\bibitem[{{Eales} {et~al.}(2012){Eales}, {Smith}, {Auld}, {Baes}, {Bendo},
  {Bianchi}, {Boselli}, {Ciesla}, {Clements}, {Cooray}, {Cortese}, {Davies},
  {De Looze}, {Galametz}, {Gear}, {Gentile}, {Gomez}, {Fritz}, {Hughes},
  {Madden}, {Magrini}, {Pohlen}, {Spinoglio}, {Verstappen}, {Vlahakis}, \&
  {Wilson}}]{Eales2012}
{Eales}, S., {Smith}, M. W.~L., {Auld}, R., {et~al.} 2012, \apj, 761, 168,
  \dodoi{10.1088/0004-637X/761/2/168}

\bibitem[{{Faisst} {et~al.}(2020){Faisst}, {Schaerer}, {Lemaux}, {Oesch},
  {Fudamoto}, {Cassata}, {B{\'e}thermin}, {Capak}, {Le F{\`e}vre}, {Silverman},
  {Yan}, {Ginolfi}, {Koekemoer}, {Morselli}, {Amor{\'\i}n}, {Bardelli},
  {Boquien}, {Brammer}, {Cimatti}, {Dessauges-Zavadsky}, {Fujimoto},
  {Gruppioni}, {Hathi}, {Hemmati}, {Ibar}, {Jones}, {Khusanova}, {Loiacono},
  {Pozzi}, {Talia}, {Tasca}, {Riechers}, {Rodighiero}, {Romano}, {Scoville},
  {Toft}, {Vallini}, {Vergani}, {Zamorani}, \& {Zucca}}]{Faisst2020}
{Faisst}, A.~L., {Schaerer}, D., {Lemaux}, B.~C., {et~al.} 2020, \apjs, 247,
  61, \dodoi{10.3847/1538-4365/ab7ccd}

\bibitem[{{Feigelson} \& {Nelson}(1985)}]{Feigelson1985}
{Feigelson}, E.~D., \& {Nelson}, P.~I. 1985, \apj, 293, 192,
  \dodoi{10.1086/163225}

\bibitem[{{Foyle} {et~al.}(2012){Foyle}, {Wilson}, {Mentuch}, {Bendo},
  {Dariush}, {Parkin}, {Pohlen}, {Sauvage}, {Smith}, {Roussel}, {Baes},
  {Boquien}, {Boselli}, {Clements}, {Cooray}, {Davies}, {Eales}, {Madden},
  {Page}, \& {Spinoglio}}]{Foyle2012}
{Foyle}, K., {Wilson}, C.~D., {Mentuch}, E., {et~al.} 2012, \mnras, 421, 2917,
  \dodoi{10.1111/j.1365-2966.2012.20520.x}

\bibitem[{{Fudamoto} {et~al.}(2022){Fudamoto}, {Inoue}, \&
  {Sugahara}}]{Fudamoto2022}
{Fudamoto}, Y., {Inoue}, A.~K., \& {Sugahara}, Y. 2022, arXiv e-prints,
  arXiv:2206.01879.
\newblock \doarXiv{2206.01879}

\bibitem[{Galliano {et~al.}(2018)Galliano, Galametz, \& Jones}]{Galliano2018}
Galliano, F., Galametz, M., \& Jones, A.~P. 2018, 56, 673,
  \dodoi{10.1146/annurev-astro-081817-051900}

\bibitem[{Genzel {et~al.}(2015)Genzel, Tacconi, Lutz, Saintonge, Berta,
  Magnelli, Combes, Garc{\'{\i}}a-Burillo, Neri, Bolatto, Contini, Lilly,
  Boissier, Boone, Bouch{\'{e}}, Bournaud, Burkert, Carollo, Colina, Cooper,
  Cox, Feruglio, Schreiber, Freundlich, Gracia-Carpio, Juneau, Kovac, Lippa,
  Naab, Salome, Renzini, Sternberg, Walter, Weiner, Weiss, \&
  Wuyts}]{Genzel2015}
Genzel, R., Tacconi, L.~J., Lutz, D., {et~al.} 2015, 800, 20,
  \dodoi{10.1088/0004-637x/800/1/20}

\bibitem[{Giovanelli {et~al.}(2005)Giovanelli, Haynes, Kent, Perillat,
  Saintonge, Brosch, Catinella, Hoffman, Stierwalt, Spekkens, Lerner, Masters,
  Momjian, Rosenberg, Springob, Boselli, Charmandaris, Darling, Davies, Lambas,
  Gavazzi, Giovanardi, Hardy, Hunt, Iovino, Karachentsev, Karachentseva,
  Koopmann, Marinoni, Minchin, Muller, Putman, Pantoja, Salzer, Scodeggio,
  Skillman, Solanes, Valotto, van Driel, \& van Zee}]{Giovanelli2005}
Giovanelli, R., Haynes, M.~P., Kent, B.~R., {et~al.} 2005, 130, 2598,
  \dodoi{10.1086/497431}

\bibitem[{{Gonz{\'a}lez-L{\'o}pez} {et~al.}(2017){Gonz{\'a}lez-L{\'o}pez},
  {Bauer}, {Aravena}, {Laporte}, {Bradley}, {Carrasco}, {Carvajal}, {Demarco},
  {Infante}, {Kneissl}, {Koekemoer}, {Mu{\~n}oz Arancibia}, {Troncoso},
  {Villard}, \& {Zitrin}}]{Gonzalez-Lopez2017}
{Gonz{\'a}lez-L{\'o}pez}, J., {Bauer}, F.~E., {Aravena}, M., {et~al.} 2017,
  \aap, 608, A138, \dodoi{10.1051/0004-6361/201730961}

\bibitem[{{Gregg} {et~al.}(2022){Gregg}, {Calzetti}, \& {Heyer}}]{Gregg2022}
{Gregg}, B., {Calzetti}, D., \& {Heyer}, M. 2022, \apj, 928, 120,
  \dodoi{10.3847/1538-4357/ac558a}

\bibitem[{Groves {et~al.}(2015)Groves, Schinnerer, Leroy, Galametz, Walter,
  Bolatto, Hunt, Dale, Calzetti, Croxall, \& Jr.}]{Groves2015}
Groves, B.~A., Schinnerer, E., Leroy, A., {et~al.} 2015, 799, 96,
  \dodoi{10.1088/0004-637x/799/1/96}

\bibitem[{{Haynes} {et~al.}(2011){Haynes}, {Giovanelli}, {Martin}, {Hess},
  {Saintonge}, {Adams}, {Hallenbeck}, {Hoffman}, {Huang}, {Kent}, {Koopmann},
  {Papastergis}, {Stierwalt}, {Balonek}, {Craig}, {Higdon}, {Kornreich},
  {Miller}, {O'Donoghue}, {Olowin}, {Rosenberg}, {Spekkens}, {Troischt}, \&
  {Wilcots}}]{Haynes2011}
{Haynes}, M.~P., {Giovanelli}, R., {Martin}, A.~M., {et~al.} 2011, \aj, 142,
  170, \dodoi{10.1088/0004-6256/142/5/170}

\bibitem[{{Helou} {et~al.}(2004){Helou}, {Roussel}, {Appleton}, {Frayer},
  {Stolovy}, {Storrie-Lombardi}, {Hurt}, {Lowrance}, {Makovoz}, {Masci},
  {Surace}, {Gordon}, {Alonso-Herrero}, {Engelbracht}, {Misselt}, {Rieke},
  {Rieke}, {Willner}, {Pahre}, {Ashby}, {Fazio}, \& {Smith}}]{Helou2004}
{Helou}, G., {Roussel}, H., {Appleton}, P., {et~al.} 2004, \apjs, 154, 253,
  \dodoi{10.1086/422640}

\bibitem[{{Hirashita} {et~al.}(2003){Hirashita}, {Buat}, \&
  {Inoue}}]{Hirashita2003}
{Hirashita}, H., {Buat}, V., \& {Inoue}, A.~K. 2003, \aap, 410, 83,
  \dodoi{10.1051/0004-6361:20031144}

\bibitem[{{Hollenbach} \& {Salpeter}(1971)}]{Hollenbach1971}
{Hollenbach}, D., \& {Salpeter}, E.~E. 1971, \apj, 163, 155,
  \dodoi{10.1086/150754}

\bibitem[{{Hollenbach} \& {Tielens}(1997)}]{Hollenbach1997}
{Hollenbach}, D.~J., \& {Tielens}, A.~G.~G.~M. 1997, \araa, 35, 179,
  \dodoi{10.1146/annurev.astro.35.1.179}

\bibitem[{{Hughes} {et~al.}(2013){Hughes}, {Cortese}, {Boselli}, {Gavazzi}, \&
  {Davies}}]{Hughes2013}
{Hughes}, T.~M., {Cortese}, L., {Boselli}, A., {Gavazzi}, G., \& {Davies},
  J.~I. 2013, \aap, 550, A115, \dodoi{10.1051/0004-6361/201218822}

\bibitem[{{Hunt} {et~al.}(2011){Hunt}, {Izotov}, {Sauvage}, \&
  {Thuan}}]{Hunt2011}
{Hunt}, L.~K., {Izotov}, Y.~I., {Sauvage}, M., \& {Thuan}, T.~X. 2011, in EAS
  Publications Series, Vol.~46, EAS Publications Series, ed. C.~{Joblin} \&
  A.~G.~G.~M. {Tielens}, 143--148, \dodoi{10.1051/eas/1146015}

\bibitem[{{Hunt} {et~al.}(2010){Hunt}, {Thuan}, {Izotov}, \&
  {Sauvage}}]{Hunt2010}
{Hunt}, L.~K., {Thuan}, T.~X., {Izotov}, Y.~I., \& {Sauvage}, M. 2010, \apj,
  712, 164, \dodoi{10.1088/0004-637X/712/1/164}

\bibitem[{{Hunt} {et~al.}(2015){Hunt}, {Draine}, {Bianchi}, {Gordon}, {Aniano},
  {Calzetti}, {Dale}, {Helou}, {Hinz}, {Kennicutt}, {Roussel}, {Wilson},
  {Bolatto}, {Boquien}, {Croxall}, {Galametz}, {Gil de Paz}, {Koda},
  {Mu{\~n}oz-Mateos}, {Sandstrom}, {Sauvage}, {Vigroux}, \&
  {Zibetti}}]{Hunt2015}
{Hunt}, L.~K., {Draine}, B.~T., {Bianchi}, S., {et~al.} 2015, \aap, 576, A33,
  \dodoi{10.1051/0004-6361/201424734}

\bibitem[{Hunter(1997)}]{Hunter1997}
Hunter, D. 1997, 109, 937, \dodoi{10.1086/133965}

\bibitem[{{Hunter}(2007)}]{Hunter2007}
{Hunter}, J.~D. 2007, Computing in Science and Engineering, 9, 90,
  \dodoi{10.1109/MCSE.2007.55}

\bibitem[{{Isobe} {et~al.}(1986){Isobe}, {Feigelson}, \& {Nelson}}]{Isobe1986}
{Isobe}, T., {Feigelson}, E.~D., \& {Nelson}, P.~I. 1986, \apj, 306, 490,
  \dodoi{10.1086/164359}

\bibitem[{Janowiecki {et~al.}(2018)Janowiecki, Cortese, Catinella, \&
  Goodwin}]{Janowiecki2018}
Janowiecki, S., Cortese, L., Catinella, B., \& Goodwin, A.~J. 2018, Monthly
  Notices of the Royal Astronomical Society, 476, 1390,
  \dodoi{10.1093/mnras/sty242}

\bibitem[{{Jarrett} {et~al.}(2011){Jarrett}, {Cohen}, {Masci}, {Wright},
  {Stern}, {Benford}, {Blain}, {Carey}, {Cutri}, {Eisenhardt}, {Lonsdale},
  {Mainzer}, {Marsh}, {Padgett}, {Petty}, {Ressler}, {Skrutskie}, {Stanford},
  {Surace}, {Tsai}, {Wheelock}, \& {Yan}}]{Jarrett2011}
{Jarrett}, T.~H., {Cohen}, M., {Masci}, F., {et~al.} 2011, \apj, 735, 112,
  \dodoi{10.1088/0004-637X/735/2/112}

\bibitem[{{Jarrett} {et~al.}(2013){Jarrett}, {Masci}, {Tsai}, {Petty},
  {Cluver}, {Assef}, {Benford}, {Blain}, {Bridge}, {Donoso}, {Eisenhardt},
  {Koribalski}, {Lake}, {Neill}, {Seibert}, {Sheth}, {Stanford}, \&
  {Wright}}]{Jarrett2013}
{Jarrett}, T.~H., {Masci}, F., {Tsai}, C.~W., {et~al.} 2013, \aj, 145, 6,
  \dodoi{10.1088/0004-6256/145/1/6}

\bibitem[{{Johnston} {et~al.}(2007){Johnston}, {Bailes}, {Bartel}, {Baugh},
  {Bietenholz}, {Blake}, {Braun}, {Brown}, {Chatterjee}, {Darling}, {Deller},
  {Dodson}, {Edwards}, {Ekers}, {Ellingsen}, {Feain}, {Gaensler}, {Haverkorn},
  {Hobbs}, {Hopkins}, {Jackson}, {James}, {Joncas}, {Kaspi}, {Kilborn},
  {Koribalski}, {Kothes}, {Landecker}, {Lenc}, {Lovell}, {Macquart},
  {Manchester}, {Matthews}, {McClure-Griffiths}, {Norris}, {Pen}, {Phillips},
  {Power}, {Protheroe}, {Sadler}, {Schmidt}, {Stairs}, {Staveley-Smith},
  {Stil}, {Taylor}, {Tingay}, {Tzioumis}, {Walker}, {Wall}, \&
  {Wolleben}}]{Johnston2007}
{Johnston}, S., {Bailes}, M., {Bartel}, N., {et~al.} 2007, \pasa, 24, 174,
  \dodoi{10.1071/AS07033}

\bibitem[{{Johnston} {et~al.}(2008){Johnston}, {Taylor}, {Bailes}, {Bartel},
  {Baugh}, {Bietenholz}, {Blake}, {Braun}, {Brown}, {Chatterjee}, {Darling},
  {Deller}, {Dodson}, {Edwards}, {Ekers}, {Ellingsen}, {Feain}, {Gaensler},
  {Haverkorn}, {Hobbs}, {Hopkins}, {Jackson}, {James}, {Joncas}, {Kaspi},
  {Kilborn}, {Koribalski}, {Kothes}, {Landecker}, {Lenc}, {Lovell}, {Macquart},
  {Manchester}, {Matthews}, {McClure-Griffiths}, {Norris}, {Pen}, {Phillips},
  {Power}, {Protheroe}, {Sadler}, {Schmidt}, {Stairs}, {Staveley-Smith},
  {Stil}, {Tingay}, {Tzioumis}, {Walker}, {Wall}, \& {Wolleben}}]{Johnston2008}
{Johnston}, S., {Taylor}, R., {Bailes}, M., {et~al.} 2008, Experimental
  Astronomy, 22, 151, \dodoi{10.1007/s10686-008-9124-7}

\bibitem[{{Jonas} \& {MeerKAT Team}(2016)}]{Jonas2016}
{Jonas}, J., \& {MeerKAT Team}. 2016, in MeerKAT Science: On the Pathway to the
  SKA, 1

\bibitem[{{Jones} {et~al.}(2013){Jones}, {Fanciullo}, {K{\"o}hler},
  {Verstraete}, {Guillet}, {Bocchio}, \& {Ysard}}]{Jones2013}
{Jones}, A.~P., {Fanciullo}, L., {K{\"o}hler}, M., {et~al.} 2013, \aap, 558,
  A62, \dodoi{10.1051/0004-6361/201321686}

\bibitem[{{Jones} {et~al.}(2017){Jones}, {K{\"o}hler}, {Ysard}, {Bocchio}, \&
  {Verstraete}}]{Jones2017}
{Jones}, A.~P., {K{\"o}hler}, M., {Ysard}, N., {Bocchio}, M., \& {Verstraete},
  L. 2017, \aap, 602, A46, \dodoi{10.1051/0004-6361/201630225}

\bibitem[{{Kauffmann} {et~al.}(2003){Kauffmann}, {Heckman}, {White}, {Charlot},
  {Tremonti}, {Brinchmann}, {Bruzual}, {Peng}, {Seibert}, {Bernardi},
  {Blanton}, {Brinkmann}, {Castander}, {Cs{\'a}bai}, {Fukugita}, {Ivezic},
  {Munn}, {Nichol}, {Padmanabhan}, {Thakar}, {Weinberg}, \&
  {York}}]{Kauffmann2003}
{Kauffmann}, G., {Heckman}, T.~M., {White}, S. D.~M., {et~al.} 2003, \mnras,
  341, 33, \dodoi{10.1046/j.1365-8711.2003.06291.x}

\bibitem[{{Kaur} {et~al.}(2022){Kaur}, {Kanekar}, {Rafelski}, {Neeleman},
  {Prochaska}, \& {Revalski}}]{Kaur2022}
{Kaur}, B., {Kanekar}, N., {Rafelski}, M., {et~al.} 2022, \apjl, 933, L42,
  \dodoi{10.3847/2041-8213/ac7bdd}

\bibitem[{{Kennicutt}(1998)}]{Kennicutt1998}
{Kennicutt}, Robert~C., J. 1998, \araa, 36, 189,
  \dodoi{10.1146/annurev.astro.36.1.189}

\bibitem[{{Kennicutt} {et~al.}(2003){Kennicutt}, {Armus}, {Bendo}, {Calzetti},
  {Dale}, {Draine}, {Engelbracht}, {Gordon}, {Grauer}, {Helou}, {Hollenbach},
  {Jarrett}, {Kewley}, {Leitherer}, {Li}, {Malhotra}, {Regan}, {Rieke},
  {Rieke}, {Roussel}, {Smith}, {Thornley}, \& {Walter}}]{Kennicutt2003}
{Kennicutt}, Robert~C., J., {Armus}, L., {Bendo}, G., {et~al.} 2003, \pasp,
  115, 928, \dodoi{10.1086/376941}

\bibitem[{{Kennicutt} {et~al.}(2011){Kennicutt}, {Calzetti}, {Aniano},
  {Appleton}, {Armus}, {Beir{\~a}o}, {Bolatto}, {Brandl}, {Crocker}, {Croxall},
  {Dale}, {Donovan Meyer}, {Draine}, {Engelbracht}, {Galametz}, {Gordon},
  {Groves}, {Hao}, {Helou}, {Hinz}, {Hunt}, {Johnson}, {Koda}, {Krause},
  {Leroy}, {Li}, {Meidt}, {Montiel}, {Murphy}, {Rahman}, {Rix}, {Roussel},
  {Sandstrom}, {Sauvage}, {Schinnerer}, {Skibba}, {Smith}, {Srinivasan},
  {Vigroux}, {Walter}, {Wilson}, {Wolfire}, \& {Zibetti}}]{Kennicutt2011}
{Kennicutt}, R.~C., {Calzetti}, D., {Aniano}, G., {et~al.} 2011, \pasp, 123,
  1347, \dodoi{10.1086/663818}

\bibitem[{{Kewley} \& {Dopita}(2002)}]{Kewley2002}
{Kewley}, L.~J., \& {Dopita}, M.~A. 2002, \apjs, 142, 35,
  \dodoi{10.1086/341326}

\bibitem[{{Kewley} {et~al.}(2001){Kewley}, {Dopita}, {Sutherland}, {Heisler},
  \& {Trevena}}]{Kewley2001}
{Kewley}, L.~J., {Dopita}, M.~A., {Sutherland}, R.~S., {Heisler}, C.~A., \&
  {Trevena}, J. 2001, \apj, 556, 121, \dodoi{10.1086/321545}

\bibitem[{{Kewley} \& {Ellison}(2008)}]{Kewley2008}
{Kewley}, L.~J., \& {Ellison}, S.~L. 2008, \apj, 681, 1183,
  \dodoi{10.1086/587500}

\bibitem[{{Kewley} {et~al.}(2019){Kewley}, {Nicholls}, \&
  {Sutherland}}]{Kewley2019}
{Kewley}, L.~J., {Nicholls}, D.~C., \& {Sutherland}, R.~S. 2019, \araa, 57,
  511, \dodoi{10.1146/annurev-astro-081817-051832}

\bibitem[{{Koribalski} {et~al.}(2020){Koribalski}, {Staveley-Smith},
  {Westmeier}, {Serra}, {Spekkens}, {Wong}, {Lee-Waddell}, {Lagos},
  {Obreschkow}, {Ryan-Weber}, {Zwaan}, {Kilborn}, {Bekiaris}, {Bekki},
  {Bigiel}, {Boselli}, {Bosma}, {Catinella}, {Chauhan}, {Cluver}, {Colless},
  {Courtois}, {Crain}, {de Blok}, {D{\'e}nes}, {Duffy}, {Elagali}, {Fluke},
  {For}, {Heald}, {Henning}, {Hess}, {Holwerda}, {Howlett}, {Jarrett}, {Jones},
  {Jones}, {J{\'o}zsa}, {Jurek}, {J{\"u}tte}, {Kamphuis}, {Karachentsev},
  {Kerp}, {Kleiner}, {Kraan-Korteweg}, {L{\'o}pez-S{\'a}nchez}, {Madrid},
  {Meyer}, {Mould}, {Murugeshan}, {Norris}, {Oh}, {Oosterloo}, {Popping},
  {Putman}, {Reynolds}, {Rhee}, {Robotham}, {Ryder}, {Schr{\"o}der}, {Shao},
  {Stevens}, {Taylor}, {van{\^A} der Hulst}, {Verdes-Montenegro}, {Wakker},
  {Wang}, {Whiting}, {Winkel}, \& {Wolf}}]{Koribalski2020}
{Koribalski}, B.~S., {Staveley-Smith}, L., {Westmeier}, T., {et~al.} 2020,
  \apss, 365, 118, \dodoi{10.1007/s10509-020-03831-4}

\bibitem[{{Krumholz}(2012)}]{Krumholz2012}
{Krumholz}, M.~R. 2012, \apj, 759, 9, \dodoi{10.1088/0004-637X/759/1/9}

\bibitem[{{Lamperti} {et~al.}(2019){Lamperti}, {Saintonge}, {De Looze},
  {Accurso}, {Clark}, {Smith}, {Wilson}, {Brinks}, {Brown}, {Bureau},
  {Clements}, {Eales}, {Glass}, {Hwang}, {Lee}, {Lin}, {Michalowski},
  {Sargent}, {Williams}, {Xiao}, \& {Yang}}]{Lamperti2019}
{Lamperti}, I., {Saintonge}, A., {De Looze}, I., {et~al.} 2019, \mnras, 489,
  4389, \dodoi{10.1093/mnras/stz2311}

\bibitem[{{Lee} {et~al.}(2013){Lee}, {Hwang}, \& {Ko}}]{Lee2013}
{Lee}, J.~C., {Hwang}, H.~S., \& {Ko}, J. 2013, \apj, 774, 62,
  \dodoi{10.1088/0004-637X/774/1/62}

\bibitem[{{Leroy} {et~al.}(2008){Leroy}, {Walter}, {Brinks}, {Bigiel}, {de
  Blok}, {Madore}, \& {Thornley}}]{Leroy2008}
{Leroy}, A.~K., {Walter}, F., {Brinks}, E., {et~al.} 2008, \aj, 136, 2782,
  \dodoi{10.1088/0004-6256/136/6/2782}

\bibitem[{Leroy {et~al.}(2011)Leroy, Bolatto, Gordon, Sandstrom, Gratier,
  Rosolowsky, Engelbracht, Mizuno, Corbelli, Fukui, \& Kawamura}]{Leroy2011}
Leroy, A.~K., Bolatto, A., Gordon, K., {et~al.} 2011, 737, 12,
  \dodoi{10.1088/0004-637x/737/1/12}

\bibitem[{{Li}(2020)}]{Li2020}
{Li}, A. 2020, Nature Astronomy, 4, 339, \dodoi{10.1038/s41550-020-1051-1}

\bibitem[{{Li} \& {Pan}(2016)}]{Li2016}
{Li}, D., \& {Pan}, Z. 2016, Radio Science, 51, 1060,
  \dodoi{10.1002/2015RS005877}

\bibitem[{{Lin} {et~al.}(2020){Lin}, {Calzetti}, {Kong}, {Adamo}, {Cignoni},
  {Cook}, {Dale}, {Grasha}, {Grebel}, {Messa}, {Sacchi}, \& {Smith}}]{Lin2020}
{Lin}, Z., {Calzetti}, D., {Kong}, X., {et~al.} 2020, \apj, 896, 16,
  \dodoi{10.3847/1538-4357/ab9106}

\bibitem[{{Madden} {et~al.}(2006){Madden}, {Galliano}, {Jones}, \&
  {Sauvage}}]{Madden2006}
{Madden}, S.~C., {Galliano}, F., {Jones}, A.~P., \& {Sauvage}, M. 2006, \aap,
  446, 877, \dodoi{10.1051/0004-6361:20053890}

\bibitem[{{Maiolino} \& {Mannucci}(2019)}]{Maiolino2019}
{Maiolino}, R., \& {Mannucci}, F. 2019, \aapr, 27, 3,
  \dodoi{10.1007/s00159-018-0112-2}

\bibitem[{{Mallory} {et~al.}(2022){Mallory}, {Calzetti}, \&
  {Lin}}]{Mallory2022}
{Mallory}, K., {Calzetti}, D., \& {Lin}, Z. 2022, \apj, 933, 156,
  \dodoi{10.3847/1538-4357/ac7227}

\bibitem[{{Martin} {et~al.}(2005){Martin}, {Fanson}, {Schiminovich},
  {Morrissey}, {Friedman}, {Barlow}, {Conrow}, {Grange}, {Jelinsky},
  {Milliard}, {Siegmund}, {Bianchi}, {Byun}, {Donas}, {Forster}, {Heckman},
  {Lee}, {Madore}, {Malina}, {Neff}, {Rich}, {Small}, {Surber}, {Szalay},
  {Welsh}, \& {Wyder}}]{Martin2005}
{Martin}, D.~C., {Fanson}, J., {Schiminovich}, D., {et~al.} 2005, \apjl, 619,
  L1, \dodoi{10.1086/426387}

\bibitem[{{Meixner} {et~al.}(2019){Meixner}, {Cooray}, {Leisawitz}, {Staguhn},
  {Armus}, {Battersby}, {Bauer}, {Bergin}, {Bradford}, {Ennico-Smith},
  {Fortney}, {Kataria}, {Melnick}, {Milam}, {Narayanan}, {Padgett},
  {Pontoppidan}, {Pope}, {Roellig}, {Sandstrom}, {Stevenson}, {Su}, {Vieira},
  {Wright}, {Zmuidzinas}, {Sheth}, {Benford}, {Mamajek}, {Neff}, {De Beck},
  {Gerin}, {Helmich}, {Sakon}, {Scott}, {Vavrek}, {Wiedner}, {Carey},
  {Burgarella}, {Moseley}, {Amatucci}, {Carter}, {DiPirro}, {Wu}, {Beaman},
  {Beltran}, {Bolognese}, {Bradley}, {Corsetti}, {D'Asto}, {Denis}, {Derkacz},
  {Earle}, {Fantano}, {Folta}, {Gavares}, {Generie}, {Hilliard}, {Howard},
  {Jamil}, {Jamison}, {Lynch}, {Martins}, {Petro}, {Ramspacher}, {Rao},
  {Sandin}, {Stoneking}, {Tompkins}, \& {Webster}}]{Meixner2019}
{Meixner}, M., {Cooray}, A., {Leisawitz}, D., {et~al.} 2019, arXiv e-prints,
  arXiv:1912.06213.
\newblock \doarXiv{1912.06213}

\bibitem[{{Morisset} {et~al.}(2016){Morisset}, {Delgado-Inglada},
  {S{\'a}nchez}, {Galbany}, {Garc{\'\i}a-Benito}, {Husemann}, {Marino}, {Mast},
  \& {Roth}}]{Morisset2016}
{Morisset}, C., {Delgado-Inglada}, G., {S{\'a}nchez}, S.~F., {et~al.} 2016,
  \aap, 594, A37, \dodoi{10.1051/0004-6361/201628559}

\bibitem[{Mu{\~{n}}oz-Mateos {et~al.}(2009)Mu{\~{n}}oz-Mateos, de~Paz,
  Boissier, Zamorano, Dale, P{\'{e}}rez-Gonz{\'{a}}lez, Gallego, Madore, Bendo,
  Thornley, Draine, Boselli, Buat, Calzetti, Moustakas, \&
  Kennicutt}]{MuozMateos2009}
Mu{\~{n}}oz-Mateos, J.~C., de~Paz, A.~G., Boissier, S., {et~al.} 2009, The
  Astrophysical Journal, 701, 1965, \dodoi{10.1088/0004-637x/701/2/1965}

\bibitem[{{Neugebauer} {et~al.}(1984){Neugebauer}, {Habing}, {van Duinen},
  {Aumann}, {Baud}, {Beichman}, {Beintema}, {Boggess}, {Clegg}, {de Jong},
  {Emerson}, {Gautier}, {Gillett}, {Harris}, {Hauser}, {Houck}, {Jennings},
  {Low}, {Marsden}, {Miley}, {Olnon}, {Pottasch}, {Raimond}, {Rowan-Robinson},
  {Soifer}, {Walker}, {Wesselius}, \& {Young}}]{Neugebauer1984}
{Neugebauer}, G., {Habing}, H.~J., {van Duinen}, R., {et~al.} 1984, \apjl, 278,
  L1, \dodoi{10.1086/184209}

\bibitem[{{Ochsenbein} {et~al.}(2000){Ochsenbein}, {Bauer}, \&
  {Marcout}}]{Ochsenbein2000}
{Ochsenbein}, F., {Bauer}, P., \& {Marcout}, J. 2000, \aaps, 143, 23,
  \dodoi{10.1051/aas:2000169}

\bibitem[{{Orellana} {et~al.}(2017){Orellana}, {Nagar}, {Elbaz},
  {Calder{\'o}n-Castillo}, {Leiton}, {Ibar}, {Magnelli}, {Daddi}, {Messias},
  {Cerulo}, \& {Slater}}]{Orellana2017}
{Orellana}, G., {Nagar}, N.~M., {Elbaz}, D., {et~al.} 2017, \aap, 602, A68,
  \dodoi{10.1051/0004-6361/201629009}

\bibitem[{{Pettini} \& {Pagel}(2004)}]{Pettini2004}
{Pettini}, M., \& {Pagel}, B. E.~J. 2004, \mnras, 348, L59,
  \dodoi{10.1111/j.1365-2966.2004.07591.x}

\bibitem[{{Pilbratt} {et~al.}(2010){Pilbratt}, {Riedinger}, {Passvogel},
  {Crone}, {Doyle}, {Gageur}, {Heras}, {Jewell}, {Metcalfe}, {Ott}, \&
  {Schmidt}}]{Pilbratt2010}
{Pilbratt}, G.~L., {Riedinger}, J.~R., {Passvogel}, T., {et~al.} 2010, \aap,
  518, L1, \dodoi{10.1051/0004-6361/201014759}

\bibitem[{{Popescu} {et~al.}(2000){Popescu}, {Misiriotis}, {Kylafis}, {Tuffs},
  \& {Fischera}}]{Popescu2000}
{Popescu}, C.~C., {Misiriotis}, A., {Kylafis}, N.~D., {Tuffs}, R.~J., \&
  {Fischera}, J. 2000, \aap, 362, 138.
\newblock \doarXiv{astro-ph/0008098}

\bibitem[{{Ramasawmy} {et~al.}(2022){Ramasawmy}, {Klaassen}, {Cicone},
  {Mroczkowski}, {Chen}, {Cornish}, {da Cunha}, {Hatziminaoglou}, {Johnstone},
  {Liu}, {Perrott}, {Schimek}, {Stanke}, \& {Wedemeyer}}]{Ramasawmy2022}
{Ramasawmy}, J., {Klaassen}, P.~D., {Cicone}, C., {et~al.} 2022, in Society of
  Photo-Optical Instrumentation Engineers (SPIE) Conference Series, Vol. 12190,
  Millimeter, Submillimeter, and Far-Infrared Detectors and Instrumentation for
  Astronomy XI, ed. J.~{Zmuidzinas} \& J.-R. {Gao}, 1219007,
  \dodoi{10.1117/12.2627505}

\bibitem[{{R{\'e}my-Ruyer} {et~al.}(2014){R{\'e}my-Ruyer}, {Madden},
  {Galliano}, {Galametz}, {Takeuchi}, {Asano}, {Zhukovska}, {Lebouteiller},
  {Cormier}, {Jones}, {Bocchio}, {Baes}, {Bendo}, {Boquien}, {Boselli},
  {DeLooze}, {Doublier-Pritchard}, {Hughes}, {Karczewski}, \&
  {Spinoglio}}]{Remy-Ruyer2014}
{R{\'e}my-Ruyer}, A., {Madden}, S.~C., {Galliano}, F., {et~al.} 2014, \aap,
  563, A31, \dodoi{10.1051/0004-6361/201322803}

\bibitem[{{R{\'e}my-Ruyer} {et~al.}(2015){R{\'e}my-Ruyer}, {Madden},
  {Galliano}, {Lebouteiller}, {Baes}, {Bendo}, {Boselli}, {Ciesla}, {Cormier},
  {Cooray}, {Cortese}, {De Looze}, {Doublier-Pritchard}, {Galametz}, {Jones},
  {Karczewski}, {Lu}, \& {Spinoglio}}]{Remy-Ruyer2015}
---. 2015, \aap, 582, A121, \dodoi{10.1051/0004-6361/201526067}

\bibitem[{{Riechers} {et~al.}(2020){Riechers}, {Boogaard}, {Decarli},
  {Gonz{\'a}lez-L{\'o}pez}, {Smail}, {Walter}, {Aravena}, {Carilli}, {Cortes},
  {Cox}, {D{\'\i}az-Santos}, {Hodge}, {Inami}, {Ivison}, {Kaasinen}, {Wagg},
  {Wei{\ss}}, \& {van der Werf}}]{Riechers2020}
{Riechers}, D.~A., {Boogaard}, L.~A., {Decarli}, R., {et~al.} 2020, \apjl, 896,
  L21, \dodoi{10.3847/2041-8213/ab9595}

\bibitem[{{Rieke} {et~al.}(2009){Rieke}, {Alonso-Herrero}, {Weiner},
  {P{\'e}rez-Gonz{\'a}lez}, {Blaylock}, {Donley}, \& {Marcillac}}]{Rieke2009}
{Rieke}, G.~H., {Alonso-Herrero}, A., {Weiner}, B.~J., {et~al.} 2009, \apj,
  692, 556, \dodoi{10.1088/0004-637X/692/1/556}

\bibitem[{{Saintonge} \& {Catinella}(2022)}]{Saintonge2022}
{Saintonge}, A., \& {Catinella}, B. 2022, arXiv e-prints, arXiv:2202.00690.
\newblock \doarXiv{2202.00690}

\bibitem[{Saintonge {et~al.}(2011)Saintonge, Kauffmann, Kramer, Tacconi,
  Buchbender, Catinella, Fabello, Graci{\'{a}}-Carpio, Wang, Cortese, Fu,
  Genzel, Giovanelli, Guo, Haynes, Heckman, Krumholz, Lemonias, Li, Moran,
  Rodriguez-Fernandez, Schiminovich, Schuster, \& Sievers}]{Saintonge2011}
Saintonge, A., Kauffmann, G., Kramer, C., {et~al.} 2011, 415, 32,
  \dodoi{10.1111/j.1365-2966.2011.18677.x}

\bibitem[{{Saintonge} {et~al.}(2012){Saintonge}, {Tacconi}, {Fabello}, {Wang},
  {Catinella}, {Genzel}, {Graci{\'a}-Carpio}, {Kramer}, {Moran}, {Heckman},
  {Schiminovich}, {Schuster}, \& {Wuyts}}]{Saintonge2012}
{Saintonge}, A., {Tacconi}, L.~J., {Fabello}, S., {et~al.} 2012, \apj, 758, 73,
  \dodoi{10.1088/0004-637X/758/2/73}

\bibitem[{Saintonge {et~al.}(2017)Saintonge, Catinella, Tacconi, Kauffmann,
  Genzel, Cortese, Dav{\'{e}}, Fletcher, Graci{\'{a}}-Carpio, Kramer, Heckman,
  Janowiecki, Lutz, Rosario, Schiminovich, Schuster, Wang, Wuyts, Borthakur,
  Lamperti, \& Roberts-Borsani}]{Saintonge2017}
Saintonge, A., Catinella, B., Tacconi, L.~J., {et~al.} 2017, 233, 22,
  \dodoi{10.3847/1538-4365/aa97e0}

\bibitem[{{Saintonge} {et~al.}(2018){Saintonge}, {Wilson}, {Xiao}, {Lin},
  {Hwang}, {Tosaki}, {Bureau}, {Cigan}, {Clark}, {Clements}, {De Looze},
  {Dharmawardena}, {Gao}, {Gear}, {Greenslade}, {Lamperti}, {Lee}, {Li},
  {Micha{\l}owski}, {Mok}, {Pan}, {Sansom}, {Sargent}, {Smith}, {Williams},
  {Yang}, {Zhu}, {Accurso}, {Barmby}, {Brinks}, {Bourne}, {Brown}, {Chung},
  {Chung}, {Cibinel}, {Coppin}, {Davies}, {Davis}, {Eales}, {Fanciullo},
  {Fang}, {Gao}, {Glass}, {Gomez}, {Greve}, {He}, {Ho}, {Huang}, {Jeong},
  {Jiang}, {Jiao}, {Kemper}, {Kim}, {Kim}, {Kim}, {Ko}, {Kong}, {Lacaille},
  {Lacey}, {Lee}, {Lee}, {Lee}, {Masters}, {Oh}, {Papadopoulos}, {Park},
  {Park}, {Parsons}, {Rowlands}, {Scicluna}, {Scudder}, {Sethuram}, {Serjeant},
  {Shao}, {Sheen}, {Shi}, {Shim}, {Smith}, {Spekkens}, {Tsai}, {Verma},
  {Urquhart}, {Violino}, {Viti}, {Wake}, {Wang}, {Wouterloot}, {Yang}, {Yim},
  {Yuan}, \& {Zheng}}]{Saintonge2018}
{Saintonge}, A., {Wilson}, C.~D., {Xiao}, T., {et~al.} 2018, \mnras, 481, 3497,
  \dodoi{10.1093/mnras/sty2499}

\bibitem[{{Salim} {et~al.}(2018){Salim}, {Boquien}, \& {Lee}}]{Salim2018}
{Salim}, S., {Boquien}, M., \& {Lee}, J.~C. 2018, \apj, 859, 11,
  \dodoi{10.3847/1538-4357/aabf3c}

\bibitem[{{Salim} {et~al.}(2016){Salim}, {Lee}, {Janowiecki}, {da Cunha},
  {Dickinson}, {Boquien}, {Burgarella}, {Salzer}, \& {Charlot}}]{Salim2016}
{Salim}, S., {Lee}, J.~C., {Janowiecki}, S., {et~al.} 2016, \apjs, 227, 2,
  \dodoi{10.3847/0067-0049/227/1/2}

\bibitem[{{Sandstrom} {et~al.}(2012){Sandstrom}, {Bolatto}, {Bot}, {Draine},
  {Ingalls}, {Israel}, {Jackson}, {Leroy}, {Li}, {Rubio}, {Simon}, {Smith},
  {Stanimirovi{\'c}}, {Tielens}, \& {van Loon}}]{Sandstrom2012}
{Sandstrom}, K.~M., {Bolatto}, A.~D., {Bot}, C., {et~al.} 2012, \apj, 744, 20,
  \dodoi{10.1088/0004-637X/744/1/20}

\bibitem[{{Schruba} {et~al.}(2012){Schruba}, {Leroy}, {Walter}, {Bigiel},
  {Brinks}, {de Blok}, {Kramer}, {Rosolowsky}, {Sandstrom}, {Schuster},
  {Usero}, {Weiss}, \& {Wiesemeyer}}]{Schruba2012}
{Schruba}, A., {Leroy}, A.~K., {Walter}, F., {et~al.} 2012, \aj, 143, 138,
  \dodoi{10.1088/0004-6256/143/6/138}

\bibitem[{{Scoville} {et~al.}(2014){Scoville}, {Aussel}, {Sheth}, {Scott},
  {Sanders}, {Ivison}, {Pope}, {Capak}, {Vanden Bout}, {Manohar}, {Kartaltepe},
  {Robertson}, \& {Lilly}}]{Scoville2014}
{Scoville}, N., {Aussel}, H., {Sheth}, K., {et~al.} 2014, \apj, 783, 84,
  \dodoi{10.1088/0004-637X/783/2/84}

\bibitem[{{Scoville} {et~al.}(2016){Scoville}, {Sheth}, {Aussel}, {Vanden
  Bout}, {Capak}, {Bongiorno}, {Casey}, {Murchikova}, {Koda},
  {{\'A}lvarez-M{\'a}rquez}, {Lee}, {Laigle}, {McCracken}, {Ilbert}, {Pope},
  {Sanders}, {Chu}, {Toft}, {Ivison}, \& {Manohar}}]{Scoville2016}
{Scoville}, N., {Sheth}, K., {Aussel}, H., {et~al.} 2016, \apj, 820, 83,
  \dodoi{10.3847/0004-637X/820/2/83}

\bibitem[{{Shivaei} {et~al.}(2017){Shivaei}, {Reddy}, {Shapley}, {Siana},
  {Kriek}, {Mobasher}, {Coil}, {Freeman}, {Sanders}, {Price}, {Azadi}, \&
  {Zick}}]{Shivaei2017}
{Shivaei}, I., {Reddy}, N.~A., {Shapley}, A.~E., {et~al.} 2017, \apj, 837, 157,
  \dodoi{10.3847/1538-4357/aa619c}

\bibitem[{{Smith} {et~al.}(2007){Smith}, {Draine}, {Dale}, {Moustakas},
  {Kennicutt}, {Helou}, {Armus}, {Roussel}, {Sheth}, {Bendo}, {Buckalew},
  {Calzetti}, {Engelbracht}, {Gordon}, {Hollenbach}, {Li}, {Malhotra},
  {Murphy}, \& {Walter}}]{Smith2007}
{Smith}, J.~D.~T., {Draine}, B.~T., {Dale}, D.~A., {et~al.} 2007, \apj, 656,
  770, \dodoi{10.1086/510549}

\bibitem[{{Smith} {et~al.}(2019){Smith}, {Clark}, {De Looze}, {Lamperti},
  {Saintonge}, {Wilson}, {Accurso}, {Brinks}, {Bureau}, {Chung}, {Cigan},
  {Clements}, {Dharmawardena}, {Fanciullo}, {Gao}, {Gao}, {Gear}, {Gomez},
  {Greenslade}, {Hwang}, {Kemper}, {Lee}, {Li}, {Lin}, {Liu}, {Moln{\'a}r},
  {Mok}, {Pan}, {Sargent}, {Scicluna}, {Smith}, {Urquhart}, {Williams}, {Xiao},
  {Yang}, \& {Zhu}}]{Smith2019}
{Smith}, M. W.~L., {Clark}, C. J.~R., {De Looze}, I., {et~al.} 2019, \mnras,
  486, 4166, \dodoi{10.1093/mnras/stz1102}

\bibitem[{{Springob} {et~al.}(2005){Springob}, {Haynes}, {Giovanelli}, \&
  {Kent}}]{Springob2005}
{Springob}, C.~M., {Haynes}, M.~P., {Giovanelli}, R., \& {Kent}, B.~R. 2005,
  \apjs, 160, 149, \dodoi{10.1086/431550}

\bibitem[{{Stark} {et~al.}(2021){Stark}, {Masters}, {Avila-Reese}, {Riffel},
  {Riffel}, {Boardman}, {Zheng}, {Weijmans}, {Dillon}, {Fielder}, {Finnegan},
  {Fofie}, {Goddy}, {Harrington}, {Pace}, {Rujopakarn}, {Samanso}, {Shamsi},
  {Sharma}, {Warrick}, {Witherspoon}, \& {Wolthuis}}]{Stark2021}
{Stark}, D.~V., {Masters}, K.~L., {Avila-Reese}, V., {et~al.} 2021, \mnras,
  503, 1345, \dodoi{10.1093/mnras/stab566}

\bibitem[{{Swaters} {et~al.}(2002){Swaters}, {van Albada}, {van der Hulst}, \&
  {Sancisi}}]{Swaters2002}
{Swaters}, R.~A., {van Albada}, T.~S., {van der Hulst}, J.~M., \& {Sancisi}, R.
  2002, \aap, 390, 829, \dodoi{10.1051/0004-6361:20011755}

\bibitem[{{Thuan} {et~al.}(2004){Thuan}, {Hibbard}, \&
  {L{\'e}vrier}}]{Thuan2004}
{Thuan}, T.~X., {Hibbard}, J.~E., \& {L{\'e}vrier}, F. 2004, \aj, 128, 617,
  \dodoi{10.1086/422431}

\bibitem[{{Tremonti} {et~al.}(2004){Tremonti}, {Heckman}, {Kauffmann},
  {Brinchmann}, {Charlot}, {White}, {Seibert}, {Peng}, {Schlegel}, {Uomoto},
  {Fukugita}, \& {Brinkmann}}]{Tremonti2004}
{Tremonti}, C.~A., {Heckman}, T.~M., {Kauffmann}, G., {et~al.} 2004, \apj, 613,
  898, \dodoi{10.1086/423264}

\bibitem[{Wang {et~al.}(2020)Wang, Catinella, Saintonge, Pan, Serra, \&
  Shao}]{Wang2020}
Wang, J., Catinella, B., Saintonge, A., {et~al.} 2020, 890, 63,
  \dodoi{10.3847/1538-4357/ab68dd}

\bibitem[{Wang {et~al.}(2016)Wang, Koribalski, Serra, van~der Hulst,
  Roychowdhury, Kamphuis, \& Chengalur}]{Wang2016}
Wang, J., Koribalski, B.~S., Serra, P., {et~al.} 2016, Monthly Notices of the
  Royal Astronomical Society, 460, 2143, \dodoi{10.1093/mnras/stw1099}

\bibitem[{Wang {et~al.}(2013)Wang, Kauffmann, J{\'{o}}zsa, Serra, van~der
  Hulst, Bigiel, Brinchmann, Verheijen, Oosterloo, Wang, Li, den Heijer, \&
  Kerp}]{Wang2013}
Wang, J., Kauffmann, G., J{\'{o}}zsa, G. I.~G., {et~al.} 2013, 433, 270,
  \dodoi{10.1093/mnras/stt722}

\bibitem[{Wang {et~al.}(2014)Wang, Fu, Aumer, Kauffmann, Jozsa, Serra,
  l.~Huang, Brinchmann, van~der Hulst, \& Bigiel}]{Wang2014}
Wang, J., Fu, J., Aumer, M., {et~al.} 2014, Monthly Notices of the Royal
  Astronomical Society, 441, 2159, \dodoi{10.1093/mnras/stu649}

\bibitem[{{Werner} {et~al.}(2004){Werner}, {Roellig}, {Low}, {Rieke}, {Rieke},
  {Hoffmann}, {Young}, {Houck}, {Brandl}, {Fazio}, {Hora}, {Gehrz}, {Helou},
  {Soifer}, {Stauffer}, {Keene}, {Eisenhardt}, {Gallagher}, {Gautier}, {Irace},
  {Lawrence}, {Simmons}, {Van Cleve}, {Jura}, {Wright}, \&
  {Cruikshank}}]{Werner2004}
{Werner}, M.~W., {Roellig}, T.~L., {Low}, F.~J., {et~al.} 2004, \apjs, 154, 1,
  \dodoi{10.1086/422992}

\bibitem[{{Westmeier} {et~al.}(2022){Westmeier}, {Deg}, {Spekkens}, {Reynolds},
  {Shen}, {Gaudet}, {Goliath}, {Huynh}, {Venkataraman}, {Lin}, {O'Beirne},
  {Catinella}, {Cortese}, {D{\'e}nes}, {Elagali}, {For}, {J{\'o}zsa},
  {Howlett}, {van der Hulst}, {Jurek}, {Kamphuis}, {Kilborn}, {Kleiner},
  {Koribalski}, {Lee-Waddell}, {Murugeshan}, {Rhee}, {Serra}, {Shao},
  {Staveley-Smith}, {Wang}, {Wong}, {Zwaan}, {Allison}, {Anderson}, {Ball},
  {Bock}, {Brodrick}, {Bunton}, {Cooray}, {Gupta}, {Hayman}, {Mahony}, {Moss},
  {Ng}, {Pearce}, {Raja}, {Roxby}, {Voronkov}, {Warhurst}, {Courtois}, \&
  {Said}}]{Westmeier2022}
{Westmeier}, T., {Deg}, N., {Spekkens}, K., {et~al.} 2022, \pasa, 39, e058,
  \dodoi{10.1017/pasa.2022.50}

\bibitem[{Wright {et~al.}(2010)Wright, Eisenhardt, Mainzer, Ressler, Cutri,
  Jarrett, Kirkpatrick, Padgett, McMillan, Skrutskie, Stanford, Cohen, Walker,
  Mather, Leisawitz, Gautier, McLean, Benford, Lonsdale, Blain, Mendez, Irace,
  Duval, Liu, Royer, Heinrichsen, Howard, Shannon, Kendall, Walsh, Larsen,
  Cardon, Schick, Schwalm, Abid, Fabinsky, Naes, \& Tsai}]{Wright2010}
Wright, E.~L., Eisenhardt, P. R.~M., Mainzer, A.~K., {et~al.} 2010, The
  Astronomical Journal, 140, 1868, \dodoi{10.1088/0004-6256/140/6/1868}

\bibitem[{{Wu} {et~al.}(2005){Wu}, {Cao}, {Hao}, {Liu}, {Wang}, {Xia}, {Deng},
  \& {Young}}]{Wu2005}
{Wu}, H., {Cao}, C., {Hao}, C.-N., {et~al.} 2005, \apjl, 632, L79,
  \dodoi{10.1086/497961}

\bibitem[{{Xie} \& {Ho}(2019)}]{Xie2019}
{Xie}, Y., \& {Ho}, L.~C. 2019, \apj, 884, 136,
  \dodoi{10.3847/1538-4357/ab4200}

\bibitem[{{Yates} {et~al.}(2021){Yates}, {Henriques}, {Fu}, {Kauffmann},
  {Thomas}, {Guo}, {White}, \& {Schady}}]{Yates2021}
{Yates}, R.~M., {Henriques}, B. M.~B., {Fu}, J., {et~al.} 2021, \mnras, 503,
  4474, \dodoi{10.1093/mnras/stab741}

\bibitem[{{Yesuf} \& {Ho}(2019)}]{Yesuf2019}
{Yesuf}, H.~M., \& {Ho}, L.~C. 2019, \apj, 884, 177,
  \dodoi{10.3847/1538-4357/ab4202}

\bibitem[{{Zavala} {et~al.}(2015){Zavala}, {Yun}, {Aretxaga}, {Hughes},
  {Wilson}, {Geach}, {Egami}, {Gurwell}, {Wilner}, {Smail}, {Blain}, {Chapman},
  {Coppin}, {Dessauges-Zavadsky}, {Edge}, {Monta{\~n}a}, {Nakajima}, {Rawle},
  {S{\'a}nchez-Arg{\"u}elles}, {Swinbank}, {Webb}, \& {Zeballos}}]{Zavala2015}
{Zavala}, J.~A., {Yun}, M.~S., {Aretxaga}, I., {et~al.} 2015, \mnras, 452,
  1140, \dodoi{10.1093/mnras/stv1351}

\bibitem[{{Zhang} \& {Ho}(2022{\natexlab{a}})}]{Zhang2022}
{Zhang}, L., \& {Ho}, L.~C. 2022{\natexlab{a}}, arXiv e-prints,
  arXiv:2212.05688.
\newblock \doarXiv{2212.05688}

\bibitem[{{Zhang} \& {Ho}(2022{\natexlab{b}})}]{Zhang2022b}
---. 2022{\natexlab{b}}, arXiv e-prints, arXiv:2212.05687.
\newblock \doarXiv{2212.05687}

\bibitem[{{Zhang} \& {Ho}(2022{\natexlab{c}})}]{Zhang2022c}
---. 2022{\natexlab{c}}, arXiv e-prints, arXiv:2212.05688.
\newblock \doarXiv{2212.05688}

\bibitem[{{Zibetti} {et~al.}(2009){Zibetti}, {Charlot}, \& {Rix}}]{Zibetti2009}
{Zibetti}, S., {Charlot}, S., \& {Rix}, H.-W. 2009, \mnras, 400, 1181,
  \dodoi{10.1111/j.1365-2966.2009.15528.x}

\end{thebibliography}
\bibliographystyle{aasjournal}
\end{document}